\documentclass{article}

\usepackage{PRIMEarxiv}

\usepackage{cite}
\usepackage{amsmath,amssymb,amsfonts}
\usepackage{algorithmic}
\usepackage{graphicx}
\usepackage{float}
\usepackage{makecell}
\usepackage{textcomp}
\usepackage[TS1,T1]{fontenc}
\usepackage{array, booktabs}
\usepackage{mathtools} 
\usepackage{lipsum} 
\usepackage{caption}
\usepackage{comment}
\usepackage{subcaption}
\usepackage{upgreek}
\usepackage[euler]{textgreek}

\usepackage{caption}
\usepackage[super]{nth}
\usepackage[utf8]{inputenc} 
\usepackage[T1]{fontenc}    
\usepackage{hyperref}       
\usepackage{url}            
\usepackage{booktabs}       
\usepackage{amsfonts}       
\usepackage{nicefrac}       
\usepackage{microtype}      
\usepackage{lipsum}
\usepackage{fancyhdr}       
\usepackage{graphicx}       
\graphicspath{{media/}}     

\title{Noise-based Correction for Electrical Impedance Tomography
}

\author{
  Kai Mason, Florencia Maurino-Alperovich, David Holder and Kirill Aristovich \\
  Department of Medical Physics and Biomedical Engineering \\
  University College London \\
  London, United Kingdom \\
  \texttt{kai.mason@ucl.ac.uk} 
}

\begin{document}

\maketitle
\begin{abstract}
\textit{Introduction.} Noisy measurements frequently cause noisy and inaccurate images in impedance imaging. No post-processing technique exists to calculate the propagation of measurement noise and use this to suppress noise in the image.  \textit{Objectives.} The objectives of this work were (1) to develop a post-processing method for noise-based correction (NBC) in impedance tomography, (2) to test whether NBC improves image quality in electrical impedance tomography (EIT), (3) to determine whether it is preferable to use correlated or uncorrelated noise for NBC, (4) to test whether NBC works with \textit{in vivo} data and (5) to test whether NBC is stable across model and perturbation geometries. \textit{Methods.} EIT was performed \textit{in silico} in a 2D homogeneous circular domain and an anatomically realistic, heterogeneous 3D human head domain for four perturbations and 25 noise levels in each case. This was validated by performing EIT for four perturbations in a circular, saline tank in 2D as well as a human head-shaped saline tank with a realistic skull-like layer in 3D. Images were assessed on the error in the weighted spatial variance (WSV) with respect to the true, target image. The effect of NBC was also tested for example \textit{in vivo} EIT data of lung ventilation in a human thorax and cortical activity in a rat brain. \textit{Results.} On visual inspection, NBC maintained or increased image quality for all perturbations and noise levels in 2D and 3D, both experimentally and \textit{in silico}. Analysis of the WSV showed that NBC significantly improved the WSV in nearly all cases. When the WSV was inferior with NBC, this was either visually imperceptible or a transformation between noisy reconstructions. For \textit{in vivo} data, NBC improved image quality in all cases and preserved the expected shape of the reconstructed perturbation. In practice, uncorrelated NBC performed better than correlated NBC and is recommended as a general-use post-processing technique in EIT. 
\end{abstract}

\section{Introduction}
\label{sec:introduction}

\subsection{Background}
\label{sec: background}

Image reconstruction in electrical impedance tomography (EIT) is performed by injecting an alternating current (AC) between pairs of electrodes through an electrically conductive region of interest (ROI) and measuring the voltage on all other electrodes. EIT can be performed in 2D or 3D in three distinct modes: absolute imaging, frequency-difference (FD) imaging and time-difference (TD) imaging. In the absolute mode, one recording of the boundary voltages is used to reconstruct an image of the interior conductivity distribution. In the FD mode, the current is injected at a spectrum of frequencies and the frequency-dependent electrical impedance of the ROI is leveraged to produce a voltage difference on the electrodes, which can be used to reconstruct an image of the materials with frequency-dependent impedance. In the TD mode, a local change in the electrical impedance over time can be measured as a difference in voltage over time which can be used to reconstruct an image of the change \cite{Holder2021}. 

 EIT is most commonly used in the TD mode for imaging lung ventilation with a single ring of electrodes \cite{Adler1997}; however, EIT has many other applications including imaging of stroke lesions in the brain (FD) \cite{Goren2018, Dowrick2016}, haemodynamic changes in the brain (TD) \cite{Hannan2021}, epileptiform activity in the brain (TD)\cite{Hannan2018, Hannan2020}, evoked neural activity in the brain (TD) \cite{Aristovich2016, Faulkner2018}, evoked neural activity in the sciatic and vagus nerves (TD) \cite{Ravagli2020, Thompson2023} and breast cancers (absolute) \cite{Murphy2017}. 

The presence of noise and the resulting low signal-to-noise ratio (SNR) in EIT recordings and images is one of the largest sources of error in EIT which severely hampers the applicability of EIT for imaging in cases when the SNR is low. The measured noise can be incorporated into the reconstruction algorithm as is done by the GREIT algorithm \cite{Adler2009, Grychtol2016}. GREIT works by defining a set of performance metrics to be tuned, optimising the reconstruction matrix (i.e. minimising the error) with a set of `training targets' and then using this reconstruction matrix for image reconstruction with new data \cite{Grychtol2016}. The benefit of this method is that the reconstruction matrix can be tailored to minimise the propagation of noise into the reconstruction and improve image quality, something which has been explicitly demonstrated to be beneficial \cite{Adler2009}. However, for cases such as fast neural EIT, there is often a large variation between models of the head, the finite element models (FEMs) themselves are generally very large (comprising 10s of millions of elements) and the shape and location of the reconstructed perturbations are not known \textit{a priori}. This means that it is likely that extensive training would be required for each model considered which is computationally intensive and time consuming.

Post-processing methods in EIT are currently dominated by machine learning techniques, including a domain-independent U-net method for improving image quality in 2D and 3D \cite{Herzberg2023}, an artificial neural network for improving image quality in 2D \cite{Martin2017} and a machine learning algorithm based on fuzzy-clustering \cite{Darma2022}. Whilst these techniques have been demonstrated to improve image quality, they do not necessarily have global application, are often uninterpretable by a human and do not seek to explicitly quantify and suppress the propagation of noise in the EIT reconstruction. This means that their application may not be global across all EIT imaging applications unless the models are comprehensively trained on a broad range of data, spanning all EIT applications.

The propagation of measurement noise in EIT has previously been considered by Frangi et al., specifically concerning the spatial distribution of noise in the imaging domain after the back-projection reconstruction algorithm was used \cite{Frangi1997, Frangi2002}. Whilst this analysis was used to study the relationship between measurement noise and image noise, this was not expanded to all linearised solutions or used as a post-processing technique to improve image quality. Noise-based correction (NBC) is a method which does exactly this (Fig. \ref{fig: NBC principle}), and it has been employed since 2017 for the application of time-difference EIT to image fast neural activity in the brain \cite{Faulkner2017, Faulkner2018}, epileptiform activity in the brain \cite{Hannan2018} and fast neural activity in nerves \cite{Ravagli2020, Thompson2023}. However, NBC has never been mathematically formalised with reference to the propagation of noise and NBC's application up to date has been based purely on heuristic interpretation of specific images. NBC has not been rigorously tested across SNRs, perturbation depths, geometries, and conductivity distributions which is a key step in ensuring that NBC is an effective, general-use post-processing tool for time-difference EIT in all cases.

\subsection{Purpose}
\label{sec: purpose}

The novelty of this work was the development of NBC as a new post-processing method for improving the quality of time-difference EIT images. This included the mathematical formalisation and justification of NBC and the testing of NBC \textit{in silico} and experimentally for EIT reconstructions in 2D and 3D with different injection protocols, perturbations, and noise. The purpose of this work was to answer the following questions:

\begin{enumerate}
    \item Is noise-based correction mathematically coherent?

    \item Does noise-based correction improve image quality in EIT?

    \item Is it preferable to use uncorrelated or correlated noise for noise-based correction?

    \item Does noise-based correction improve image quality with \textit{in vivo} EIT data?

    \item Is noise-based correction robust for a range of domain geometries and electrode positions?

\end{enumerate}

\subsection{Experimental Design}
\label{sec: experimental design}

For 2D image reconstruction, all `skip' injection protocols were considered since it is common and realistic for any of the considered skip patterns to be used experimentally \cite{LuppiSilva2017}. When performing EIT on an irregular geometry in 3D, the placement of the electrodes does not allow for clear `skip' patterns so an optimal injection protocol was used which maximised the current density in the ROI \cite{Faulkner2017}. Since this was an objectively optimised protocol used experimentally \cite{Faulkner2018}, only this protocol was considered in 3D. 

Both uncorrelated and correlated noise were considered in this study because both are significant sources of noise in experimental EIT \cite{Mason2023b}. Uncorrelated noise is equivalent to additive, background noise which was considered to be of equal amplitude on all measurement electrodes. Correlated noise is equivalent to current source noise and is proportional to the amplitude of the voltage on each electrode. 

A post-processing threshold of 50 \% the largest absolute change was applied to all images. This involved setting the reconstructed conductivity change to zero for a pixel/voxel if the value was less than one-half of the largest absolute conductivity change in the image. If NBC was used, then this threshold was applied after NBC. This was performed because it is a well-established, easily implemented and commonly used method in EIT to remove low-intensity artefacts and increase the interpretability of the images. The absolute impedance change was used to threshold instead of the signed change since, in general, the sign of the conductivity change may not be known \cite{Faulkner2018, Adler1996, Tidswell2001, Wan2010, Hannan2018}.

For the 2D and 3D simulations and tank studies, four perturbations were chosen to cover the full range of perturbation depths in the imaging domain and determine whether the effectiveness of NBC was dependent on perturbation depth \cite{Mason2023b}. For example, the perturbations in the 3D human head model were chosen to represent activity in the cortex, cingulate gyrus, thalamus and pons \cite{Mason2023b}. These range from the surface of the brain to deep brain structures respectively.

EIT images in 3D are generally of lower quality and the measurements have a lower SNR than those in 2D. Therefore, AC was injected for a longer time in 3D (10 s per pair) than in 2D (1 s per pair) to ensure successful reconstruction of images in 3D.

The weighted spatial variance (WSV) was chosen as the figure of merit for the reconstructed images because it is an objective and exhaustive criterion that takes into account spatial resolution, artefact reconstruction, amplitude response, position error, and shape deformation in 2D or 3D \cite{Javaherian2013}. 

0$^{\text{th}}$ order Tikhonov regularisation was the only reconstruction algorithm considered because, for many applications of EIT including brain imaging and three-dimensional thoracic imaging, large meshes comprising millions of elements are required to accurately model the system. For such meshes, the computational implementation of  0$^{\text{th}}$ order Tikhonov regularisation using generalised cross-validation is considerably less computationally expensive than other methods such as total variation regularisation or NOSER which would not be feasible for large meshes \cite{Gonzalez2016, Zhou2015, Cheney1990}. For this reason, only 0$^{\text{th}}$ order Tikhonov regularisation was considered; however, the mathematics of NBC hold for all reconstruction algorithms considering the linearised form of the EIT inverse problem.
\section{Noise Propagation and Noise-based Correction}
\label{sec: noise propagation}

The inverse problem of EIT is to solve the equation
\begin{equation}
    \label{eq: inverse problem}
    \textbf{v} = \textbf{F} (\textbf{\textsigma})
\end{equation}
for \textbf{\textsigma}, where \textbf{v} is the voltages on the electrodes, \textbf{\textsigma} is the conductivity in the ROI and \textbf{F} is the forward operator. The problem is nonlinear, ill-posed and ill-conditioned so is often linearised and regularised to find a solution \cite{Holder2021, Lionheart2004}. The linear form of \eqref{eq: inverse problem} equates \textbf{F(\textsigma)} with \textbf{J\textsigma}, where \textbf{J} is the Jacobian, which relates changes in conductivity in the ROI to changes in voltage on the electrodes. The FEM is used to discretise the ROI and allows \textbf{\textsigma} to be defined as constant on each element of the FEM \cite{Holder2021}. 

The propagation of noise in medical imaging can be studied by analysing the probability density function (PDF) of the noise. In general, for a random variable $X$ the PDF of $Y = f(X)$ can be expressed as 

\begin{equation}
    \label{eq: probability density function}
    \text{pdf}_Y(Y) = \frac{\text{pdf}_X(X)}{\left| f'(X) \right|}
\end{equation}
where $\text{pdf}_Y(Y)$ and $\text{pdf}_X(X)$ are the probability density functions for $Y$ and $X$ respectively \cite{Papoulis1965, Frangi1997}. 

For the case of the linearised EIT inverse problem
\begin{subequations}
    \begin{equation}
        Y= \sigma_i, \hspace{2mm} \sigma_i \in \textbf{\textsigma}, \hspace{2mm} i \in \{1,2,...,N_e\}
    \end{equation}

    \begin{equation}
        X = v_j, \hspace{2mm} v_j \in \textbf{v}, \hspace{2mm} j \in \{1,2,...,N_m\}
    \end{equation}

    \begin{equation}
         \left| f'(X) \right| = \left| \textbf{J}_{i,j}^{-1} \right|
    \end{equation}
\end{subequations}
where $N_e$ is the number of elements in the FEM, $N_m$ is the number of measurements and each $\sigma_i$ can be calculated as
\begin{equation}
    \sigma_i = \text{row}_i(\textbf{J}^{-1})\textbf{v} = \sum_{j=1}^{N_m}\textbf{J}_{i,j}^{-1}v_j  = \sum_{j=1}^{N_m}\sigma_{i_j}.
\end{equation}
 If the PDF of the noisy voltage measurement $v_j$ on each electrode is Gaussian $\mathcal{N}(\mu,\,s^{2})$ with mean $\mu$ and standard deviation $s$, then the PDF of each $\sigma_{i_{j}}$ is
\begin{equation}
\label{eq: gaussian v}
    \text{pdf}_{\sigma_{i_j}}(\sigma_{i_j}) = \frac{\mathcal{N}(\mu,\,s^{2})}{\left|\textbf{J}_{i,j}^{-1} \right|}
\end{equation}
which is a Gaussian distribution, scaled by  $\left|\textbf{J}_{i,j}^{-1} \right|$. 
The PDF of $\sigma_i$ can be found by convolving the PDFs for each $\sigma_{i_j}$ \cite{Frangi1997}
\begin{equation}
    \label{eq: pdf convolution} 
    \text{pdf}_{\sigma_i(\sigma_{i})} = \text{pdf}_{\sigma_{i_1}}(\sigma_{i_1})*\text{pdf}_{\sigma_{i_2}}(\sigma_{i_2})*...*\text{pdf}_{\sigma_{i_{N_{m}}}}(\sigma_{i_{N_m}})
\end{equation} 
which is known to produce a scaled Gaussian distribution \cite{Bromiley2003}. Thus, it can be shown, that for the linearised form of the inverse problem of EIT, Gaussian noise on the measurement electrodes will propagate as Gaussian noise in the reconstructed conductivity distribution.


\subsection{Principle of Noise-based Correction}
\begin{figure}[H]
    \centering
    \includegraphics[width = 0.8\textwidth]{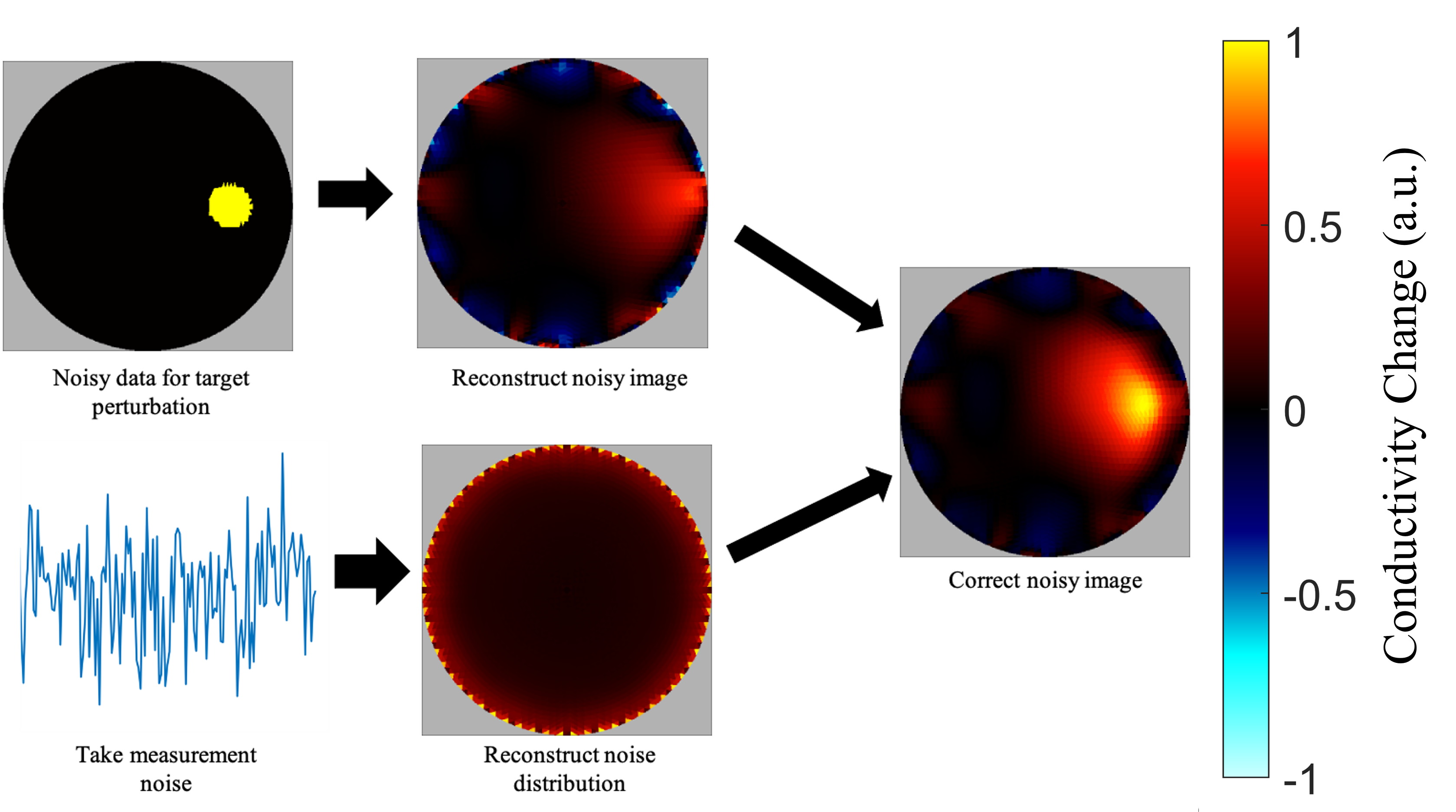}
    \caption{The principle of noise-based correction demonstrated for a 2D image.}
    \label{fig: NBC principle}
\end{figure}

NBC is a post-processing method for noise reduction in impedance imaging. This is achieved by calculating the amplitude of noise in the reconstructed image and then suppressing pixels/voxels of the image based on their respective noise. It has been shown that if the noise in the data is Gaussian, then the reconstruction process will preserve the Gaussian form, meaning the noise can be characterised by its standard deviation $s$, representing the amplitude of the noise. In order to calculate a correction factor for a problem with $N_s$ samples, $N_m$ measurement channels and $N_e$ elements in the FEM, the noise $\textbf{N} \in \mathbb{R}^{N_m\times N_s}$ can be reconstructed to calculate $\textbf{M} \in \mathbb{R}^{N_e \times N_s}$ as 

\begin{equation}
\label{eq: M}
    \textbf{M} = \textbf{J}^{-1}\textbf{N}
\end{equation}
and the standard deviation of \textbf{M}, $ s_{\textbf{M}} \in \mathbb{R}^{N_e \times 1}$ can be calculated as a correction factor (Fig. \ref{fig: NBC principle}). The noise-corrected conductivity can then be expressed as 

\begin{equation}
    \label{eq: NBC}
    (\sigma_{\text{NBC}})_i = \frac{(\sigma_0)_i}{(s_\textbf{M})_i}
\end{equation}
where $\sigma_0$ is the uncorrected solution and $i \in \{1,2,...,N_e\}$ represents the $i^{\text{th}}$ component.

When performing NBC, the noise used can either be uncorrelated, simulated noise which is the same amplitude on all measurement channels, or it can be the real/measured noise which is likely of a different amplitude on different measurement channels (i.e. correlated noise). The former will henceforth be referred to as `uncorrelated NBC (U-NBC)' and the latter as `correlated NBC (C-NBC)'.
\section{Methods}

All simulations in this study were performed using the EIDORS software in Matlab \cite{Polydorides2002}.

\subsection{2D Computational Model}
The forward problem was solved on a 200 mm diameter circular 2D finite element model (FEM) comprising 6400 triangular elements and 32 boundary electrodes equally spaced on the circumference of the model. Four perturbations of 15 mm diameter and 1 \% change in conductivity were considered at depths of 15.0 mm, 43.3 mm, 71.7 mm and 100 mm from the edge of the tank measured to the centre of mass of the perturbation. Correlated and uncorrelated artificial, zero-mean Gaussian noise were separately added to the simulated voltages. Uncorrelated noise was considered at 25 amplitudes (logarithmically spaced) between 1 nV and 1 \textmu V and correlated noise was considered at 25 amplitudes (logarithmically spaced) between 0.001 \% and 1 \% of the standing voltages. 

The images were reconstructed on a FEM representing the same geometry comprising 4096 triangular elements. The alternating current (AC) was injected at 1 mA using 15 different `skip-$N$' injection protocols, where $N \in \{0, 2, ..., 14\}$  (i.e. skip-0 is between electrodes [1,2], [2,3], ..., [32,1]). 100 reconstructions were performed for uncorrelated and correlated noise using \nth{0} order Tikhonov Regularisation. U-NBC was performed for uncorrelated noise and both U-NBC and C-NBC were performed for correlated noise.

\subsection{3D computational Model}

The forward problem was solved on an anatomically realistic model of a human head comprising seven tissue types and 3.2 M tetrahedral elements \cite{Mason2023b}. A hexahedral mesh of 375 k cubic elements was used for image reconstruction. 32 scalp electrodes of 5 mm radius and 1 k\textOmega \hspace{1mm} contact impedance were considered in the EEG 10-20 positions on the scalp of the model and a ground node was considered at the nape of the model for single-ended measurement. The injection protocol was calculated such that the current density was maximised in the brain (grey and white matter) of the model and 1 mA was the injection amplitude. Four approximately spherical perturbations of 1 \% local conductivity change were considered at depths of 7.40 mm, 32.8 mm, 58.2 mm and 83.6 mm from the inside of the skull to the centre of the perturbation. The volume of each perturbation was kept constant at 3.86 cm$^3$ and only included elements corresponding to grey or white matter. The model and perturbations used were identical to those in \cite{Mason2023b}. 

Correlated and uncorrelated artificial, zero-mean Gaussian noise were separately added to the simulated voltages. Uncorrelated noise was considered at 25 amplitudes (logarithmically spaced) between 1 nV and 10 \textmu V and correlated noise was considered at 25 amplitudes (logarithmically spaced) between 0.00001 \% and 0.1 \% of the standing voltages. 100 reconstructions were performed for uncorrelated and correlated noise using \nth{0} order Tikhonov Regularisation. U-NBC was performed for uncorrelated noise and both U-NBC and C-NBC were performed for correlated noise.

\subsection{2D Tank Experiment}
The experiment was performed on a tank of 200 mm diameter with 32 stainless steel electrodes of 3 mm radius equally spaced about the perimeter and one additional reference electrode. The tank was filled with 0.2 \% saline and AC was injected at 1475 Hz and 1 mA peak for 1 s per injection pair. The experiment was repeated for injection protocols from `skip-0' to `skip-14' and the voltage on each electrode was measured at a sampling frequency of 50 kHz. A cylindrical resistive plastic perturbation of 30 mm diameter was imaged at depths of 15 mm, 43.3 mm, 71.7 mm and 100 mm from the edge of the tank as measured to the centre of mass of the perturbation. Images were reconstructed on a circular FEM comprising 4096 triangular elements using \nth{0} order Tikhonov regularisation without NBC, with U-NBC and C-NBC. The data for each injection was separated into 100 separate epochs which were individually used to reconstruct an image. 

\subsection{3D Tank Experiment}
A tank was 3D-printed in the geometry of an adult human head truncated at the level of the forehead with a synthetic skull of accurate conductivity distribution, this was identical to the system developed by Avery et al. \cite{Avery2017}. 33 stainless steel electrodes of 5 mm radius were placed at the EEG 10-20 positions on the model boundary \cite{Avery2017}. 32 electrodes were used for AC injection and measurement and one was used as a reference electrode. Four spherical 3D-printed plastic perturbations of 20 mm diameter were imaged at depths (from the inside of the skull to the centre of mass of the perturbation) of 10 mm, 40 mm, 70 mm and 100 mm from the inside of the skull. The tank was filled with 0.2 \% saline and AC was injected at 1475 Hz and 1 mA peak for 10 s per injection pair, the voltage on each electrode was measured at a sampling frequency of 50 kHz. The injection protocol was chosen to maximise the current density in the saline (i.e. not in the skull) \cite{Faulkner2017}. The data for each injection were separated into 100 separate epochs and individually reconstructed on a hexahedral FEM comprising 253 k elements using \nth{0} order Tikhonov regularisation without NBC, with U-NBC and C-NBC.

\subsection{In vivo Lung Data}
Lung ventilation data was collected and image reconstruction was performed with and without NBC. One ring of 32 electrodes was placed around the thorax of a healthy subject who was instructed to take slow, deep breaths. Current was injected at 5 kHz and 1 mA in a `skip-10' pattern. 

Images were reconstructed using 0$^{\text{th}}$ order Tikhonov regularisation and U-NBC and C-NBC were applied separately. The noise was calculated as the standard deviation of the voltage during a period of breath-hold. Images were reconstructed in 3D on a FEM of a human thorax (conductivity = 1 S/m) comprising 26 k elements using EIDORS \cite{Polydorides2002}. Since no true reference image was available, the images were reconstructed and assessed heuristically based on visual inspection and prior knowledge and expectations of the physiological phenomena. It was expected that two, lung-shaped regions of decreased conductivity would be visible upon inhalation.

\subsection{Noise-based Correction with \textit{in vivo} Fast Neural EIT Data}

Previously collected data for imaging of fast neural activity were obtained and image reconstruction was performed with and without NBC. 

The fast neural imaging data was collected by Faulkner et al. \cite{Faulkner2019} using an array of 47 electrodes on the cortex of a rat with the ScouseTom EIT system \cite{Avery2017b}. The whisker of the rat was stimulated at 1 Hz and EIT data was collected by injecting current between pairs of electrodes at 1475 Hz and 50 \textmu A using a protocol which optimised the current density in the ROI \cite{Faulkner2017, Faulkner2018, Faulkner2019}.

Images were reconstructed using 0$^{\text{th}}$ order Tikhonov regularisation and U-NBC and C-NBC were applied separately. The noise was calculated as the standard deviation of the voltage during the baseline period (i.e. no evoked potential was present) and images were reconstructed in 3D on a FEM of a rat brain comprising three tissue types corresponding to grey matter (conductivity = 0.3 S/m), white matter (conductivity = 0.15 S/m) and cerebrospinal fluid (conductivity = 0.7 S/m). Since no true reference image was available, the images were reconstructed and assessed heuristically based on visual inspection and prior knowledge and expectations of the physiological phenomena. For neural activity due to whisker stimulation in the rat, it was expected that a region of increased conductivity within the barrel cortex of the rat would be visible \cite{Faulkner2018, Faulkner2019}. 

\subsection{EIT System}
The ScouseTom EIT system was used for both tank experiments and the lung ventilation data \cite{Avery2017b}. This comprised a Keithley constant current source \cite{Keithley}, a commercial amplifier \cite{BrainProducts2016}, a 24-bit analogue-to-digital converter \cite{BrainProducts2016} and a custom micro-controller and multiplexer \cite{Avery2017b}. The intrinsic correlated noise of the system is that of the Keithley 6221 current source which is 0.25 \% \cite{Keithley}; the uncorrelated noise is environment dependent but is expected to be $\sim 5$ \textmu V \cite{Mason2023}.

\subsection{Figure of Merit}
The images were thresholded at 50 \% of the maximum absolute change and the image quality was assessed using the weighted spatial variance (WSV) \cite{Javaherian2013} defined as

\begin{equation}
    WSV = \left(\sum_{i=1}^{n}w_i((x_i - \bar{x})^2 + (y_i - \bar{y})^2 +(z_i - \bar{z}^2)) \right)^{\frac{1}{2}}
    \label{eq: WSV}
\end{equation}
where $n$ is the total number of elements in the FEM whose reconstructed conductivity change exceeds 50 \% of the largest absolute change, $x_i$, $y_i$, and $z_i$ are the coordinates of the centre of the $i^{\text{th}}$ element, $\bar{x}$, $\bar{y}$ and $\bar{z}$ are the coordinates of the centre of mass of the true perturbation and $w_i$ is a weighting for each element defined as
\begin{equation}
    w_i = \frac{S_iI_i^2}{\sum_{i=1}^{n}S_iI_i^2}
\end{equation}
where $S_i$ is the volume (3D) or area (2D) of the $i^{\text{th}}$ element and $I_i$ is the reconstructed conductivity change of the $i^{\text{th}}$ element \cite{Javaherian2013}.

The WSV for the target image $WSV_{t}$ was calculated in each case and scaled such that $WSV_{t} = 1$. This scaling factor was applied to the WSV of each reconstruction $WSV_{r}$ accordingly such that a $WSV_{r}$ closer to $1$ indicated a higher quality reconstruction. 

\subsection{Data Analysis}

The SNR was defined as the amplitude of the voltage change between the perturbed and unperturbed cases divided by the standard deviation of the voltage over the measurement time of the unperturbed case.

The median WSV was calculated when averaging across reconstructions and this was plotted as the median $\pm$ interquartile range (IQR) for the purposes of visualisation. When C-NBC and U-NBC were used, the Friedman test was used to assess statistical significance between the three groups and the Tukey method was used for pairwise comparisons. When only U-NBC was used, the Wilcoxon signed-rank test was used to test for statistical significance between the WSV with U-NBC and no NBC.

\section{Results}
\label{sec: results}

\begin{table}[H]
\footnotesize
    \centering
    \begin{tabular}{c c c c c c c}
    \hline
    \makecell{Computational \\ Model} & \makecell{Noise \\ Type} & \makecell{NBC \\ Type} & \makecell{Number \\ Improved} & \makecell{Number \\ Deteriorated} & \makecell{Largest WSV \\ Improvement} & \makecell{SNR of \\ Largest WSV  \\ Improvement} \\
    \hline
     2D & Uncorrelated & Uncorrelated & 90 & 7 & 6.1 & 2.0 \\
     2D & Correlated & Uncorrelated & 92 & 6 & 6.2 & 1.2 \\
     2D & Correlated & Correlated & 97 & 2 & 6.6 & 1.2 \\
     3D & Uncorrelated & Uncorrelated & 100 & 0 & 4.8 & 1.3 \\
     3D & Correlated & Uncorrelated & 90 & 10 & 6.0 & 1.3 \\
     3D & Correlated & Correlated & 90 & 10 & 6.6 & 1.3 \\
     \hline
    \end{tabular}
    \caption{The number, magnitude and SNR of statistically significant improvements and deteriorations in the WSV for each computational model, noise type and NBC type across all perturbations and SNRs. In each case, 100 SNRs were considered and statistically insignificant differences in the WSV were omitted from the table. An improvement in WSV indicates that the WSV of the image with NBC was numerically closer to the WSV of the target image than the WSV of the image without NBC.}
    \label{tab: WSV improvement}
\end{table}

\subsection{2D Computational Model}

\begin{figure}
    \centering
    \begin{subfigure}{0.49\textwidth}
        \centering         
        \includegraphics[width=\textwidth]{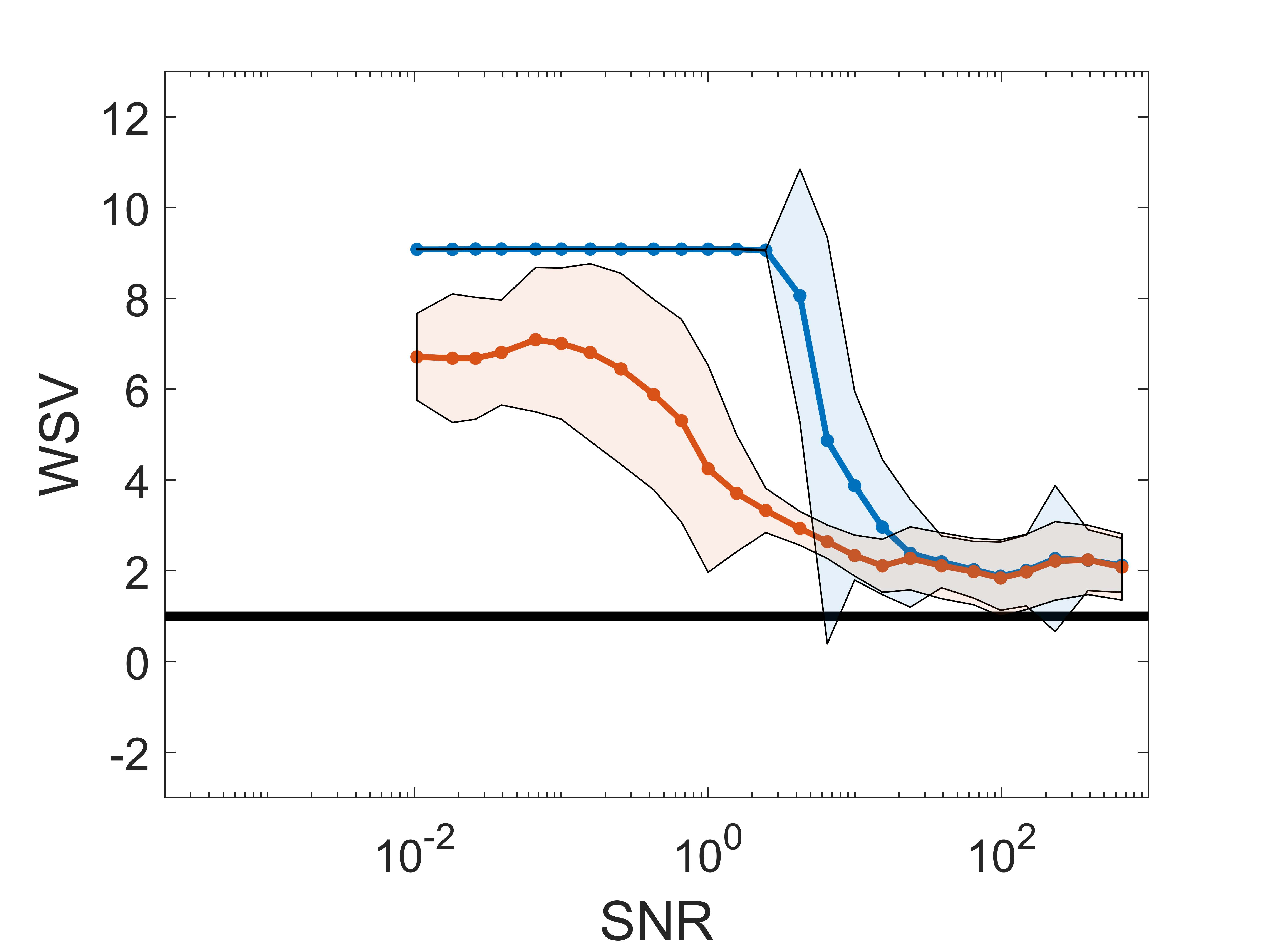}
        \captionsetup{width=0.9\linewidth}
        \caption{Perturbation 1, depth = 15.0 mm.}
        \label{fig: uncorrelated WSV 2d simulated pert 1}
    \end{subfigure}
    \hfill
    \begin{subfigure}{0.49\textwidth}
        \centering         
        \includegraphics[width=\textwidth]{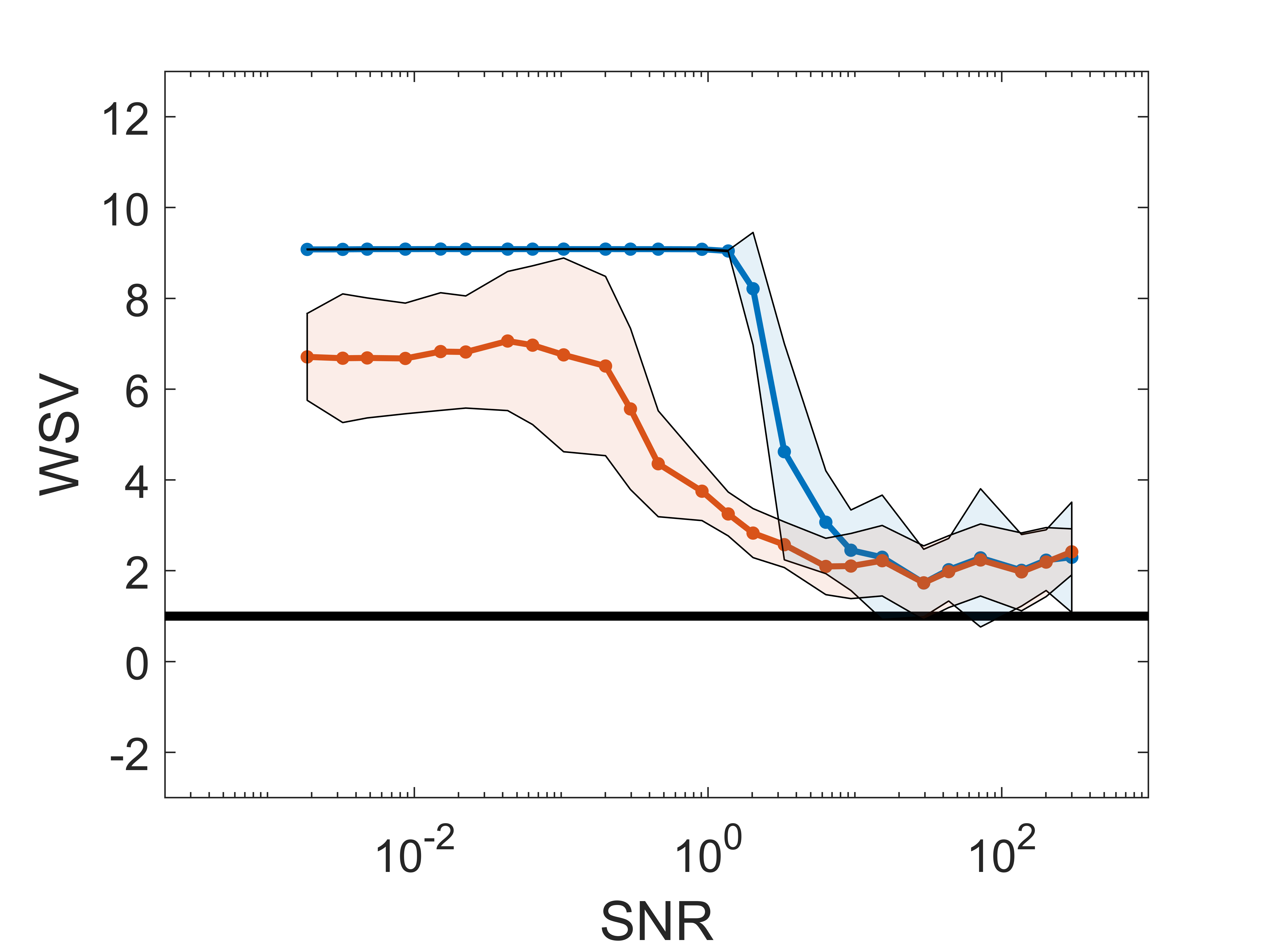}
        \captionsetup{width=0.9\linewidth}
        \caption{Perturbation 2, depth = 43.3 mm.}
        \label{fig: uncorrelated WSV 2d simulated pert 2}
    \end{subfigure}
    \hfill
    \begin{subfigure}{0.49\textwidth}
        \centering         
        \includegraphics[width=\textwidth]{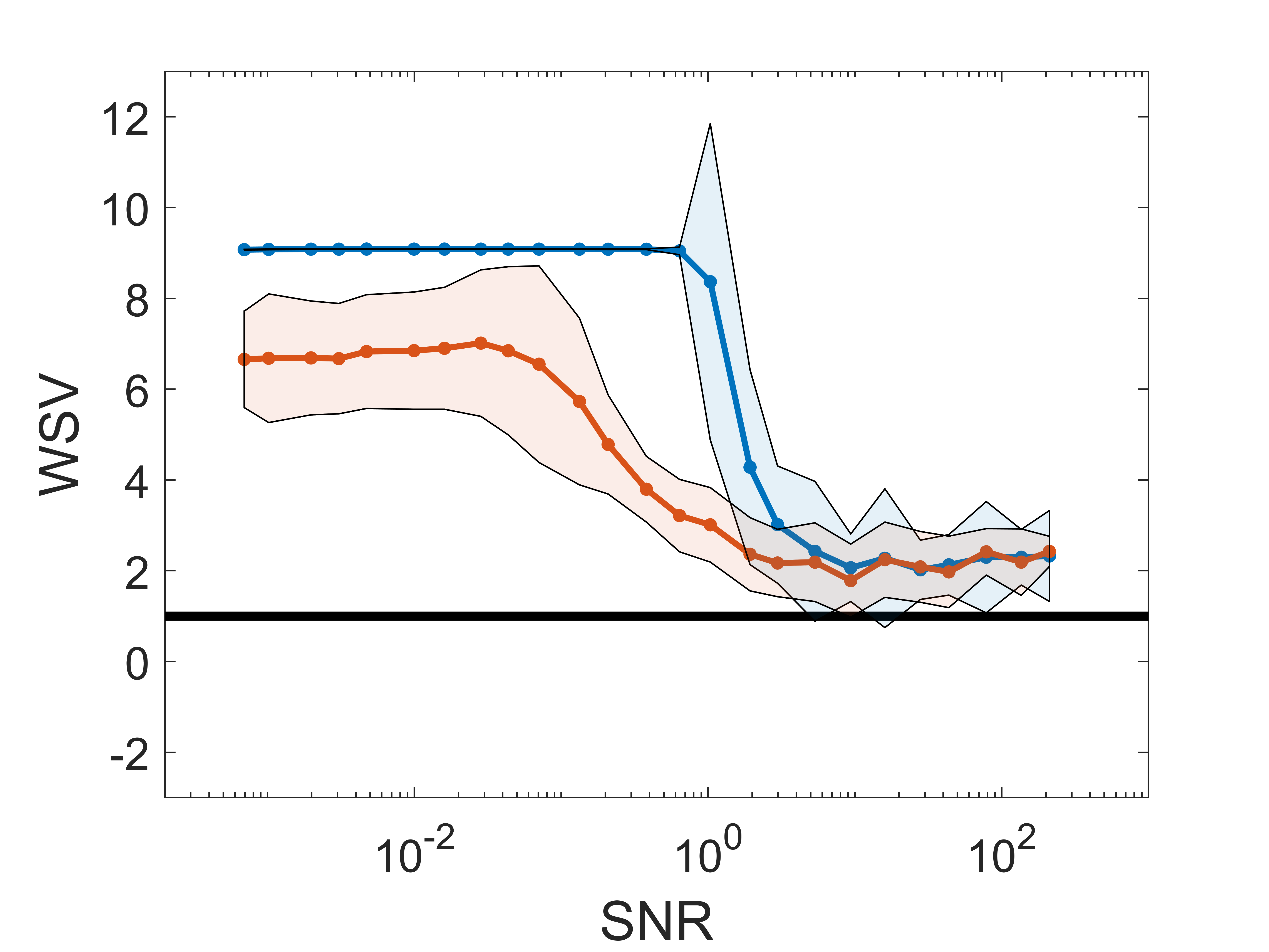}
        \captionsetup{width=0.9\linewidth}
        \caption{Perturbation 3, depth = 71.7 mm.}
        \label{fig: uncorrelated WSV 2d simulated pert 3}
    \end{subfigure}
    \hfill
    \begin{subfigure}{0.49\textwidth}
        \centering         
        \includegraphics[width=\textwidth]{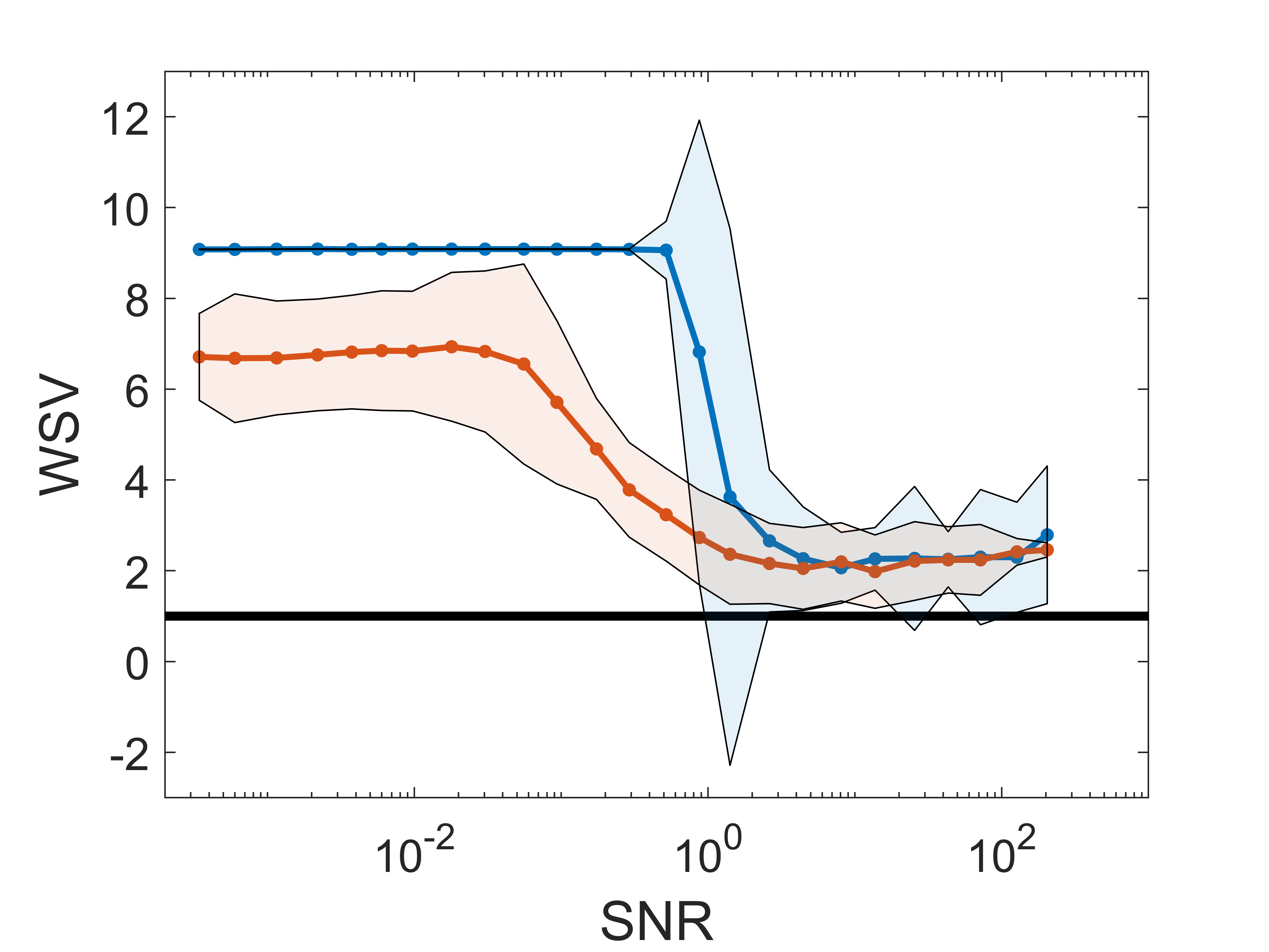}
        \captionsetup{width=0.9\linewidth}
        \caption{Perturbation 4, depth = 100 mm.}
        \label{fig: uncorrelated WSV 2d simulated pert 4}
    \end{subfigure}

    \caption{The scaled WSV (mean $\pm$ IQR) for each perturbation as a function of the SNR for images reconstructed from 2D simulated data with uncorrelated noise. The WSV is shown for images reconstructed without NBC (blue), with U-NBC (orange) and for the target image (black horizontal line). 375 SNRs were considered in total, with 100 reconstructions at each SNR. These were logarithmically binned into 25 bins. A WSV value closer to the black line indicates a higher image quality.}
    \label{fig: uncorrelated WSV 2d simulated}
\end{figure}

\begin{figure}
    \centering
    \begin{subfigure}{0.49\textwidth}
        \centering         
        \includegraphics[width=\textwidth]{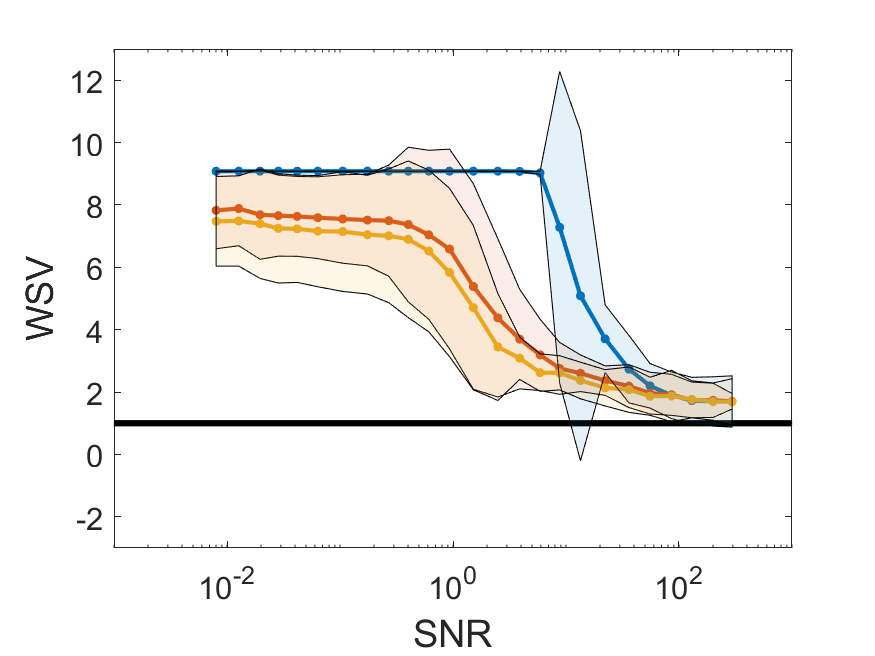}
        \captionsetup{width=0.7\linewidth}
        \caption{Perturbation 1, depth = 15.0 mm.}
        \label{fig: correlated WSV 2d simulated pert 1}
    \end{subfigure}
    \hfill
    \begin{subfigure}{0.49\textwidth}
        \centering         
        \includegraphics[width=\textwidth]{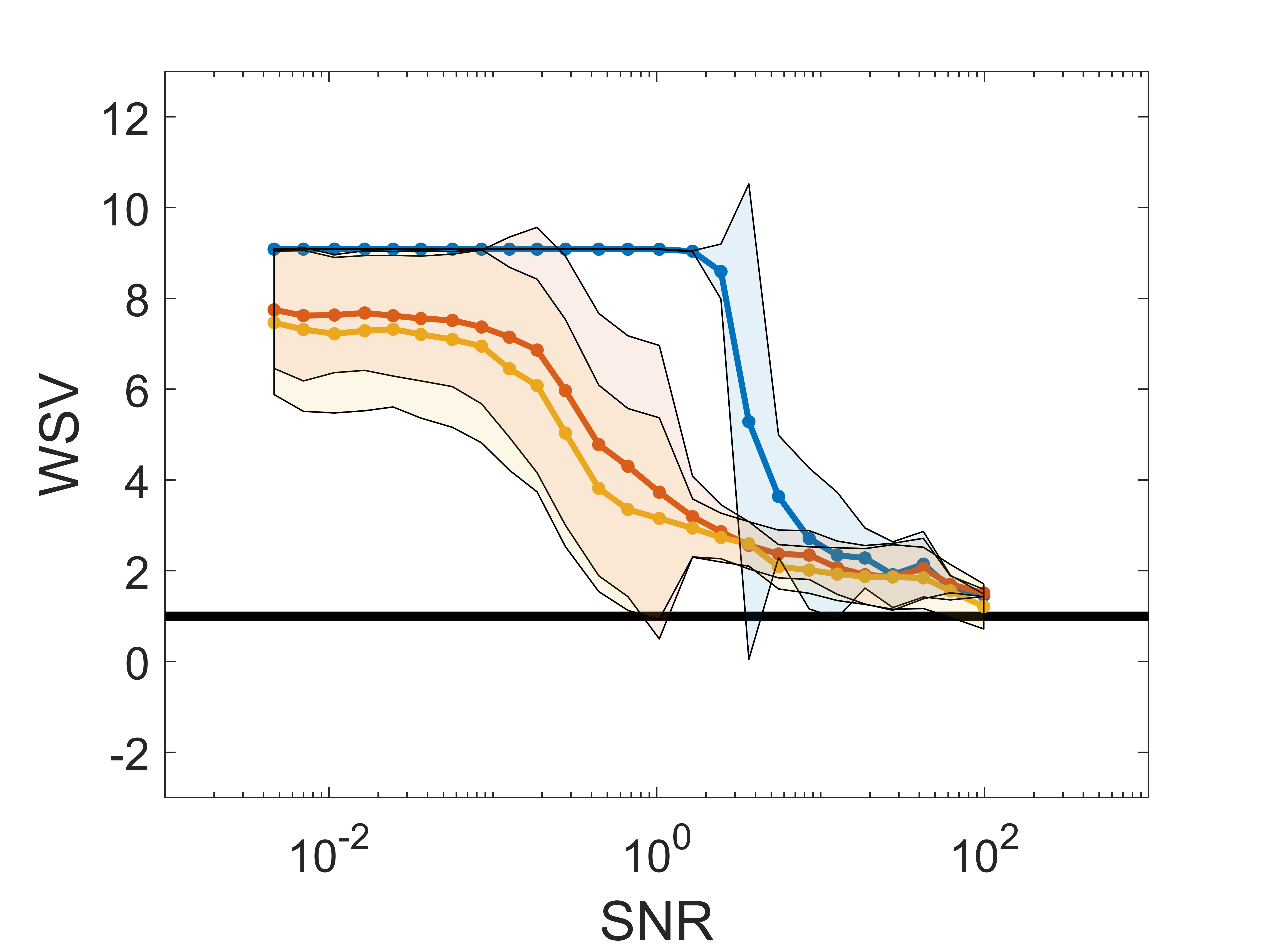}
        \captionsetup{width=0.7\linewidth}
        \caption{Perturbation 2, depth = 43.3 mm.}
        \label{fig: correlated WSV 2d simulated pert 2}
    \end{subfigure}
    \hfill
    \begin{subfigure}{0.49\textwidth}
        \centering         
        \includegraphics[width=\textwidth]{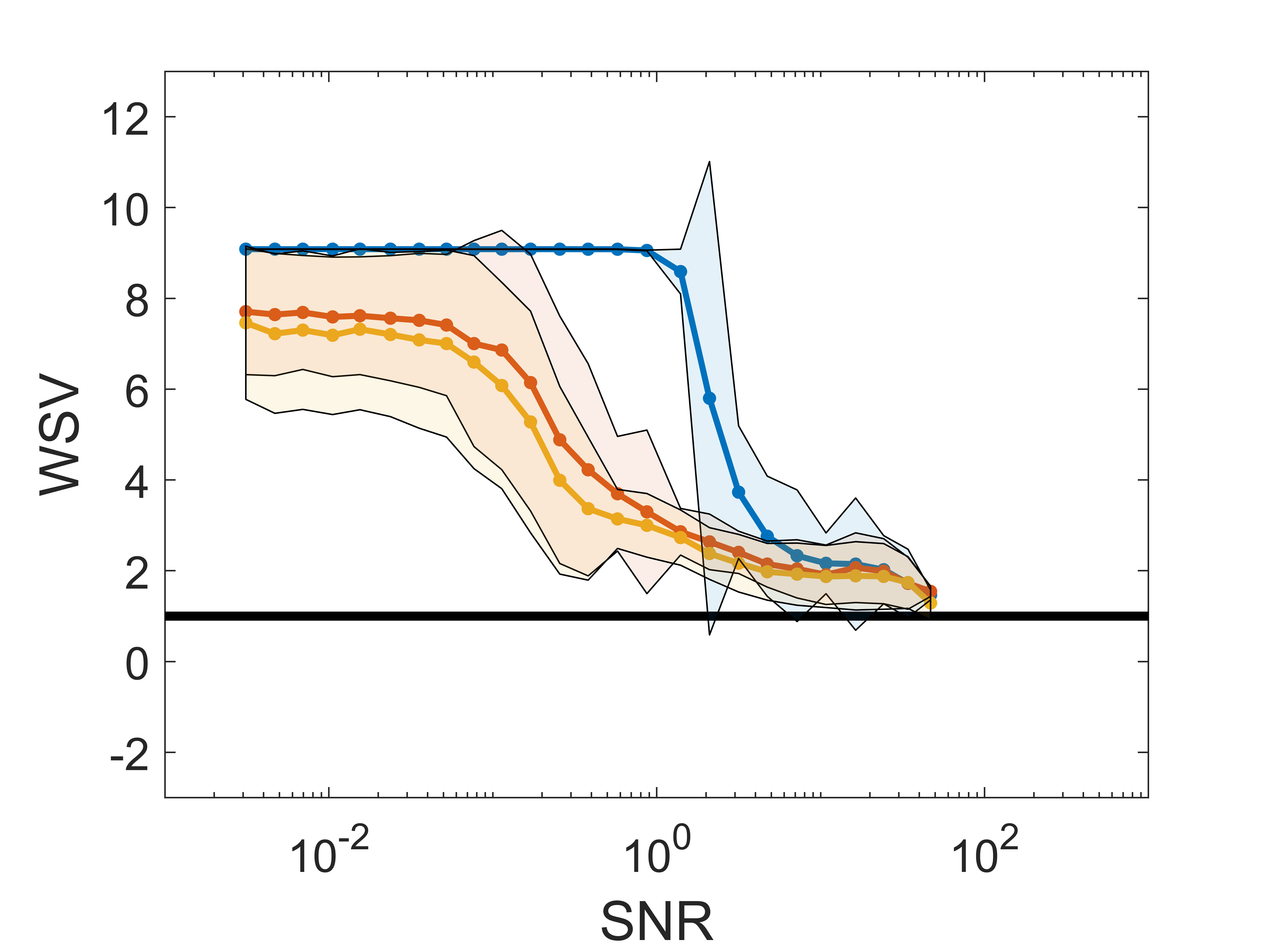}
        \captionsetup{width=0.7\linewidth}
        \caption{Perturbation 3, depth = 71.7 mm.}
        \label{fig: correlated WSV 2d simulated pert 3}
    \end{subfigure}
    \hfill
    \begin{subfigure}{0.49\textwidth}
        \centering         
        \includegraphics[width=\textwidth]{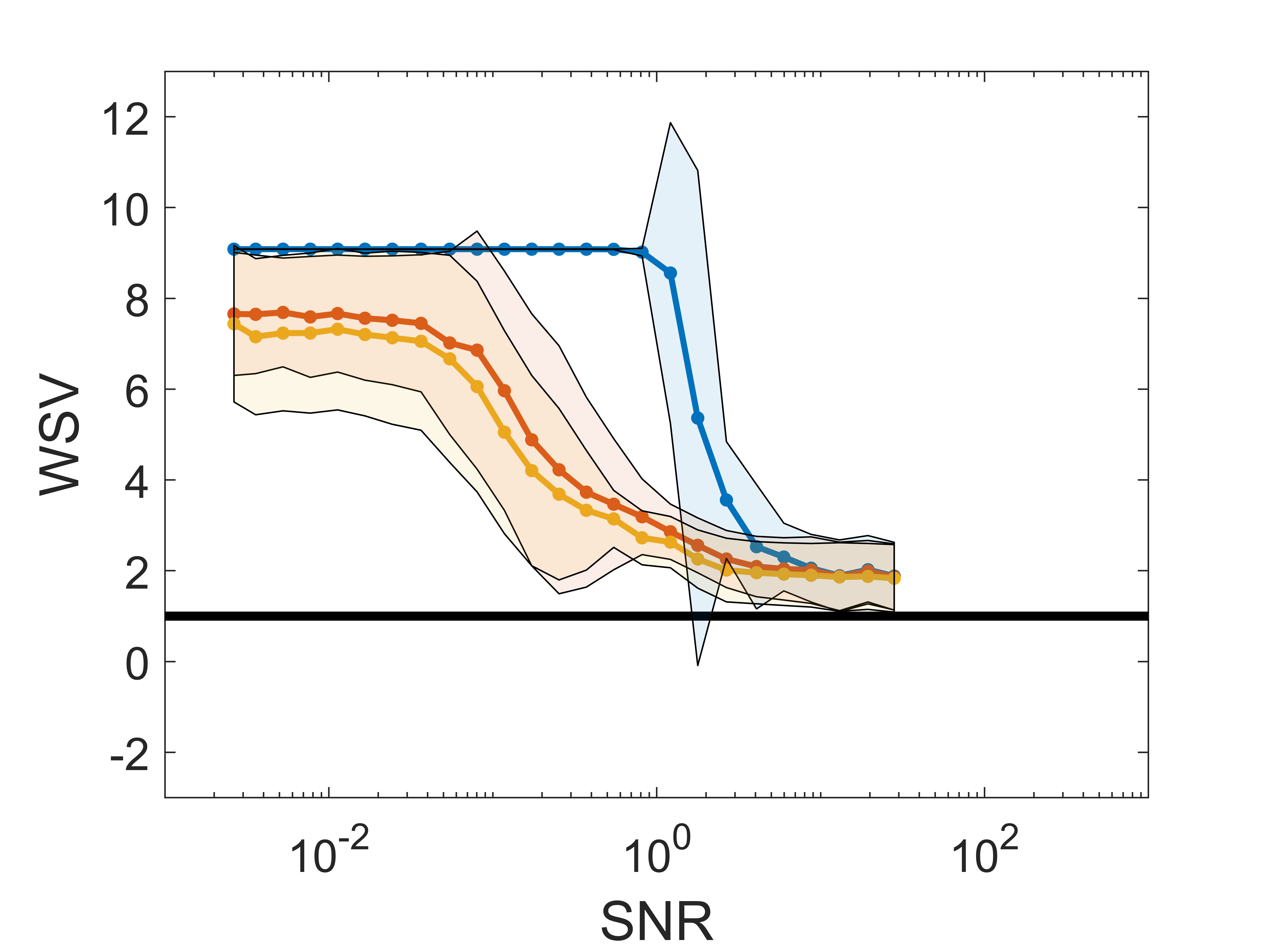}
        \captionsetup{width=0.7\linewidth}
        \caption{Perturbation 4, depth = 100 mm.}
        \label{fig: correlated WSV 2d simulated pert 4}
    \end{subfigure}

    \caption{The scaled WSV (mean $\pm$ IQR) for each perturbation as a function of the SNR for images reconstructed from 2D simulated data with correlated noise. The WSV is shown for images reconstructed without NBC (blue), with U-NBC (orange), with C-NBC (yellow) and for the target image (black horizontal line). 375 SNRs were considered in total, with 100 reconstructions at each SNR. These were logarithmically binned into 25 bins. A WSV value closer to the black line indicates a higher image quality.}
    \label{fig: correlated WSV 2d simulated}
\end{figure}

On visual inspection, reconstructions without NBC, with U-NBC and C-NBC corresponded with the target image with a fidelity that degraded with decreasing SNR. For some SNRs, images without NBC did not correspond to the target image but images with NBC did. In general, images with NBC were of equal or higher quality than those without NBC(Fig. \ref{fig: 2D simulated examples}). U-NBC and C-NBC improved the WSV for the majority of images in all noise cases (Table \ref{tab: WSV improvement} and Fig. \ref{fig: uncorrelated WSV 2d simulated} and \ref{fig: correlated WSV 2d simulated}).

\begin{figure*}
    \centering
    \begin{subfigure}{\textwidth}
        \centering
        \includegraphics[width=0.8\textwidth]{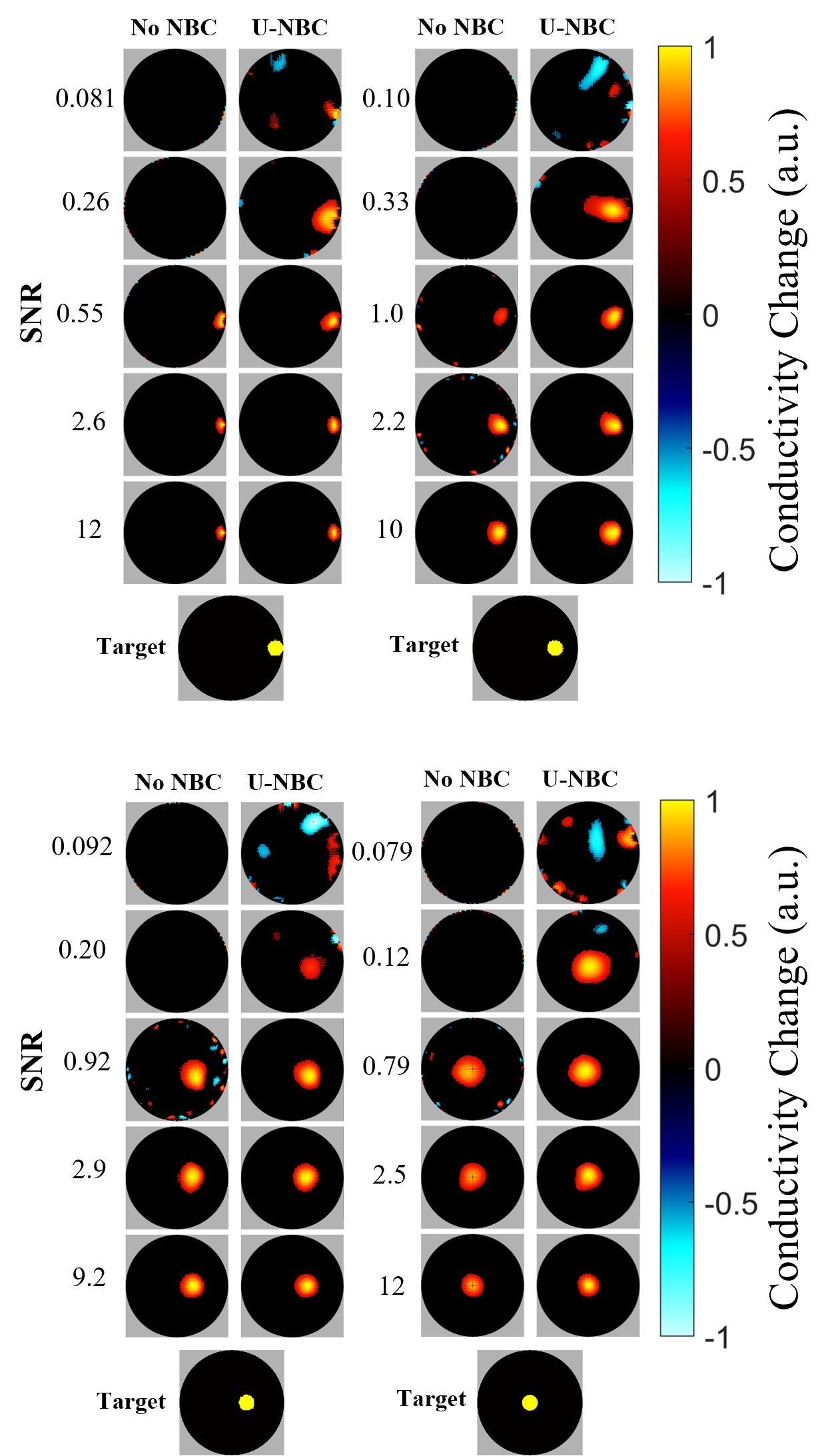}
        \caption{}
        \label{fig: uncorrelated 2D simulated example}
    \end{subfigure}
    \end{figure*}
    \begin{figure*}\ContinuedFloat
    \begin{subfigure}{\textwidth}
        \centering
        \includegraphics[width=\textwidth]{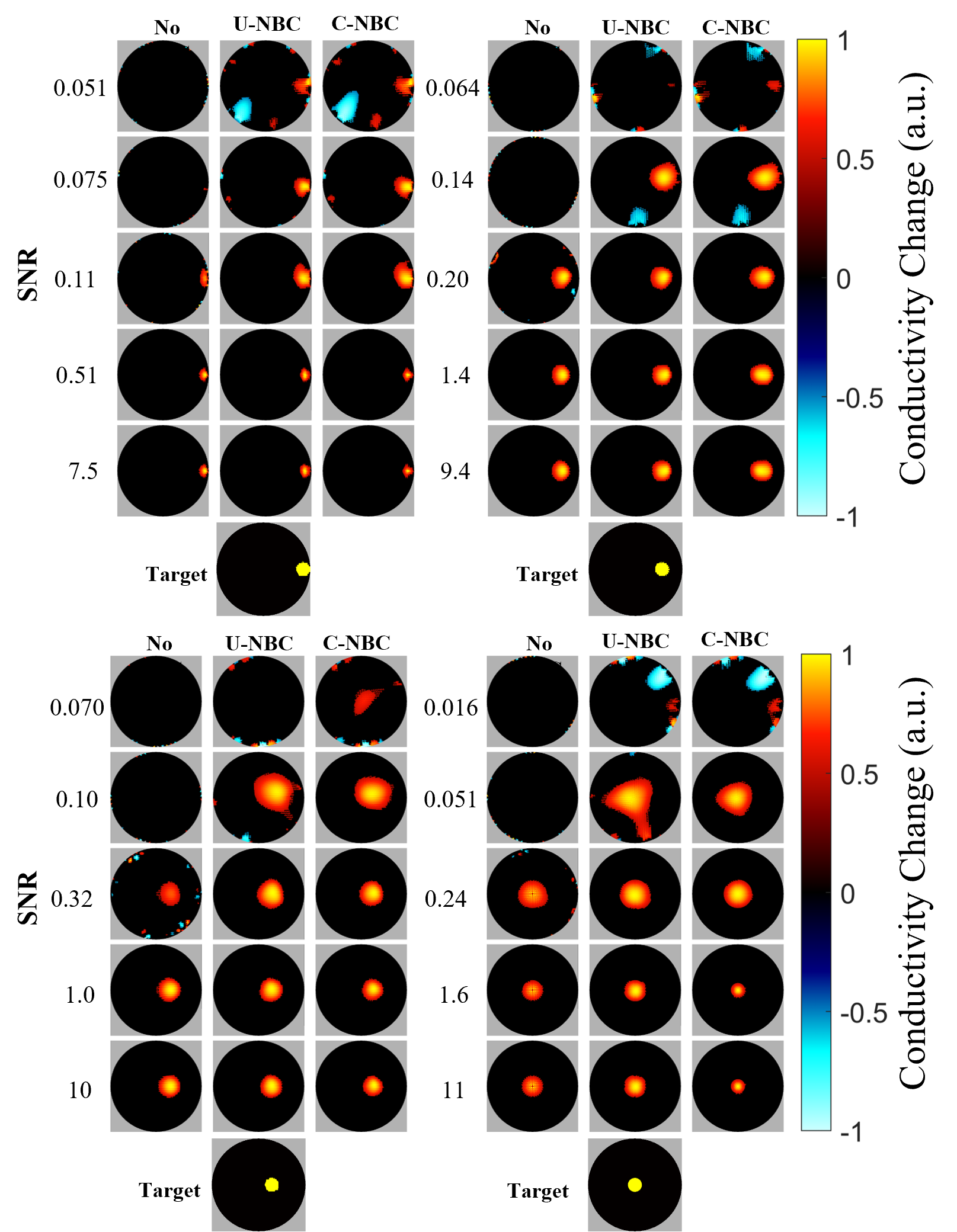}
        \caption{ }
        \label{fig: correlated 2D simulated example}
    \end{subfigure}
    \caption{Example reconstructions at five SNRs of all perturbations reconstructed from simulated data in 2D with (a) uncorrelated Gaussian noise and (b) Correlated Gaussian noise. All images were thresholded at 50 \% of the largest absolute change. A `skip-1' injection protocol was used for these images.}
    \label{fig: 2D simulated examples}
\end{figure*}

\subsection{3D Computational Model}
\begin{figure}
    \centering
    \begin{subfigure}{0.49\textwidth}
        \centering         
        \includegraphics[width=\textwidth]{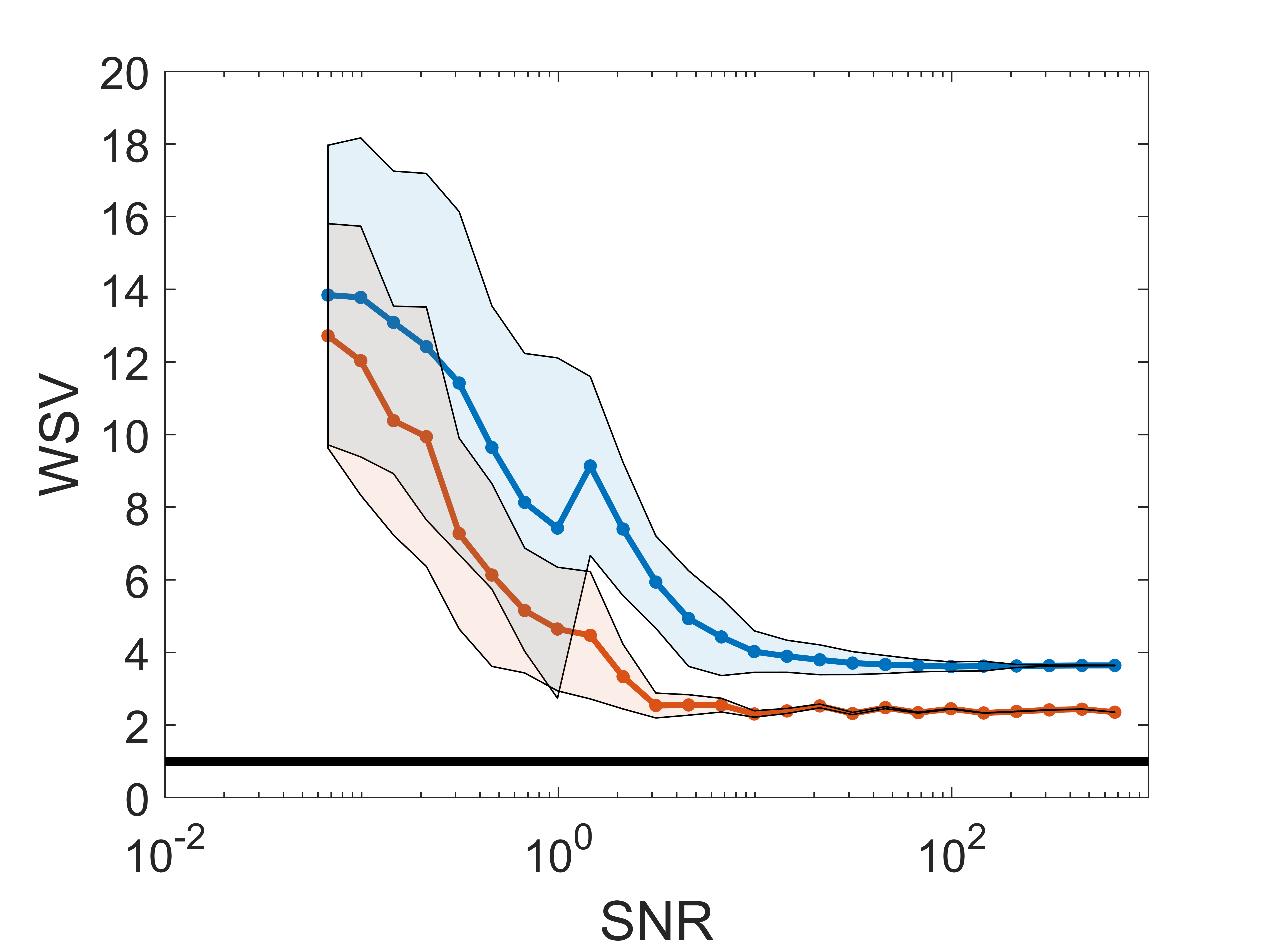}
        \captionsetup{width=0.9\linewidth}
        \caption{Perturbation 1, depth = 7.40 mm.}
        \label{fig: uncorrelated WSV 3d simulated pert 1}
    \end{subfigure}
    \hfill
    \begin{subfigure}{0.49\textwidth}
        \centering         
        \includegraphics[width=\textwidth]{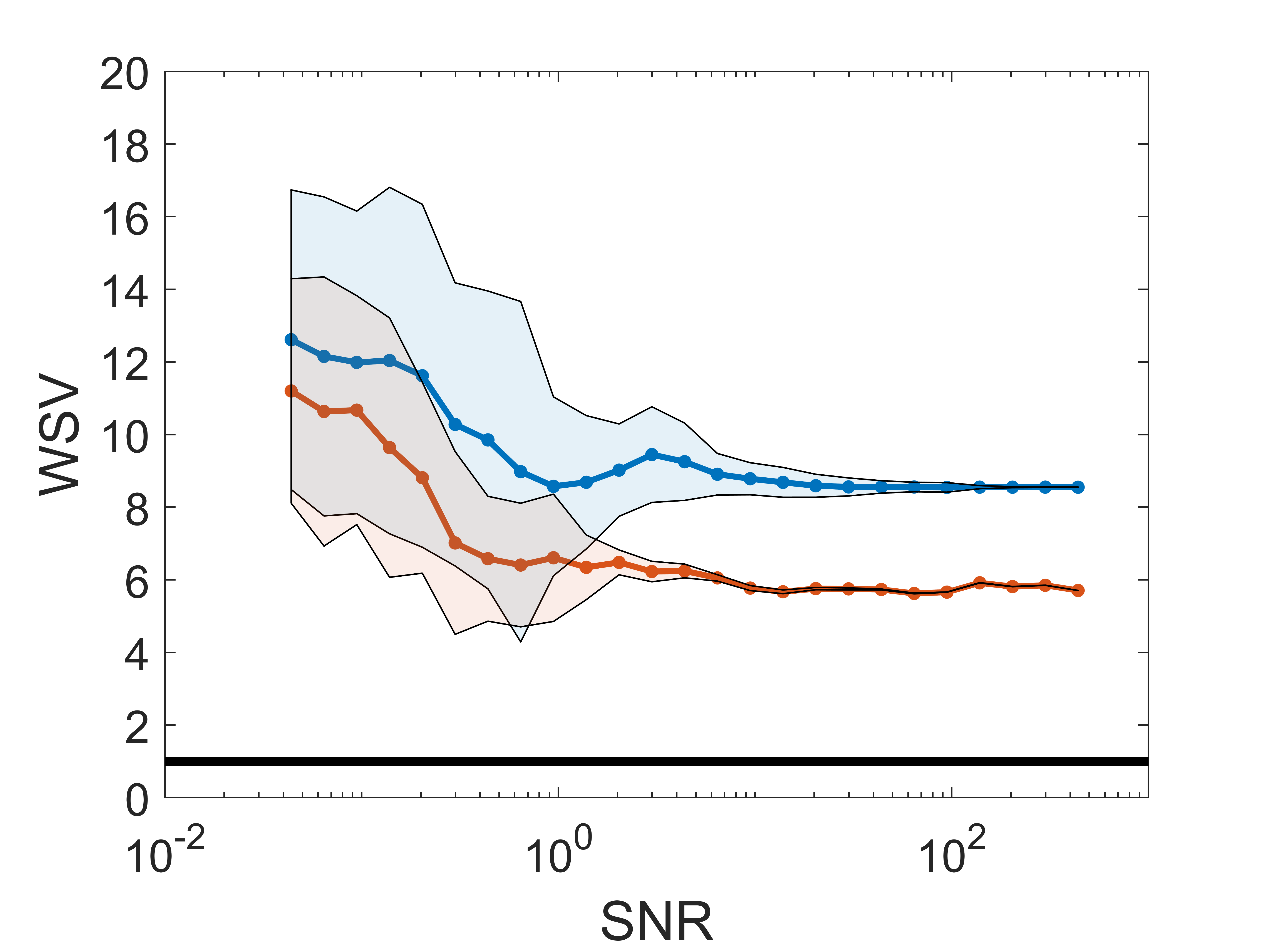}
        \captionsetup{width=0.9\linewidth}
        \caption{Perturbation 2, depth = 32.8 mm.}
        \label{fig: uncorrelated WSV 3d simulated pert 2}
    \end{subfigure}
    \hfill
    \begin{subfigure}{0.49\textwidth}
        \centering         
        \includegraphics[width=\textwidth]{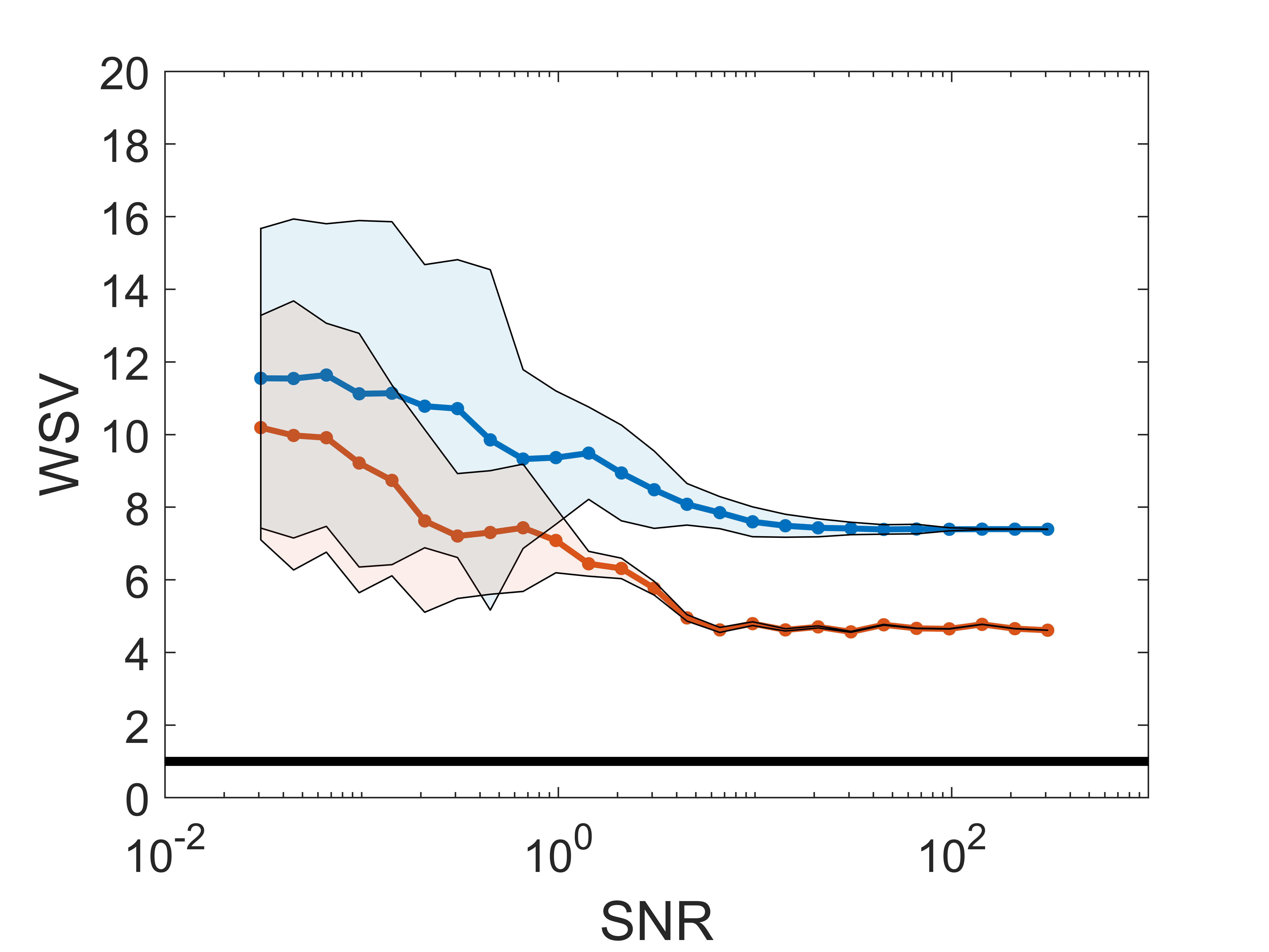}
        \captionsetup{width=0.9\linewidth}
        \caption{Perturbation 3, depth = 58.2 mm.}
        \label{fig: uncorrelated WSV 3d simulated pert 3}
    \end{subfigure}
    \hfill
    \begin{subfigure}{0.49\textwidth}
        \centering         
        \includegraphics[width=\textwidth]{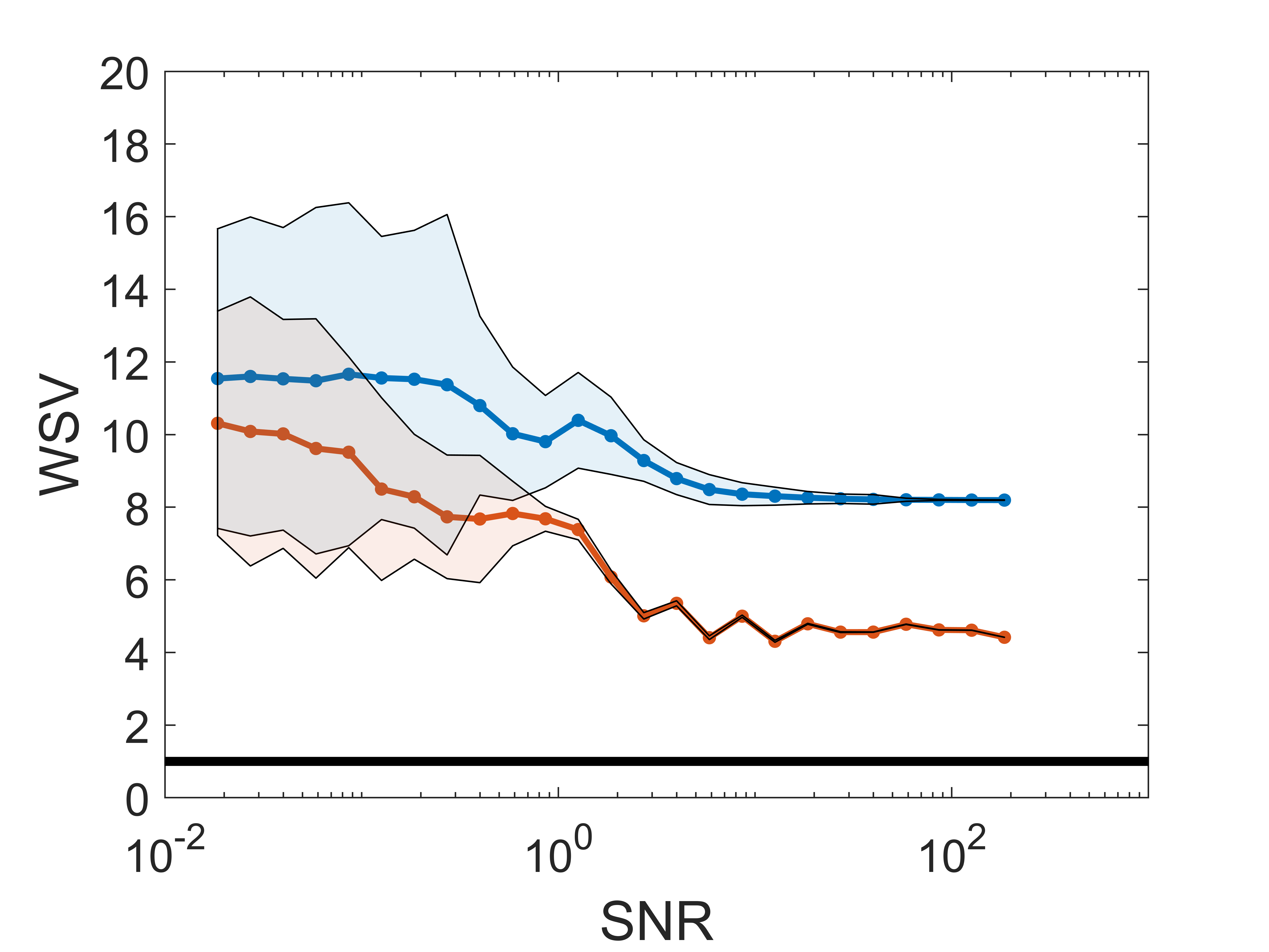}
        \captionsetup{width=0.9\linewidth}
        \caption{Perturbation 4, depth = 83.6 mm.}
        \label{fig: uncorrelated WSV 3d simulated pert 4}
    \end{subfigure}

    \caption{The scaled WSV (mean $\pm$ IQR) for each perturbation as a function of the SNR for images reconstructed from 3D simulated data with uncorrelated noise. The WSV is shown for images reconstructed without NBC (blue), with U-NBC (orange) and for the target image (black horizontal line). A WSV value closer to the black line indicates a higher image quality.}
    \label{fig: uncorrelated WSV 3d simulated}
\end{figure}

\begin{figure}
    \centering
    \begin{subfigure}{0.49\textwidth}
        \centering         
        \includegraphics[width=\textwidth]{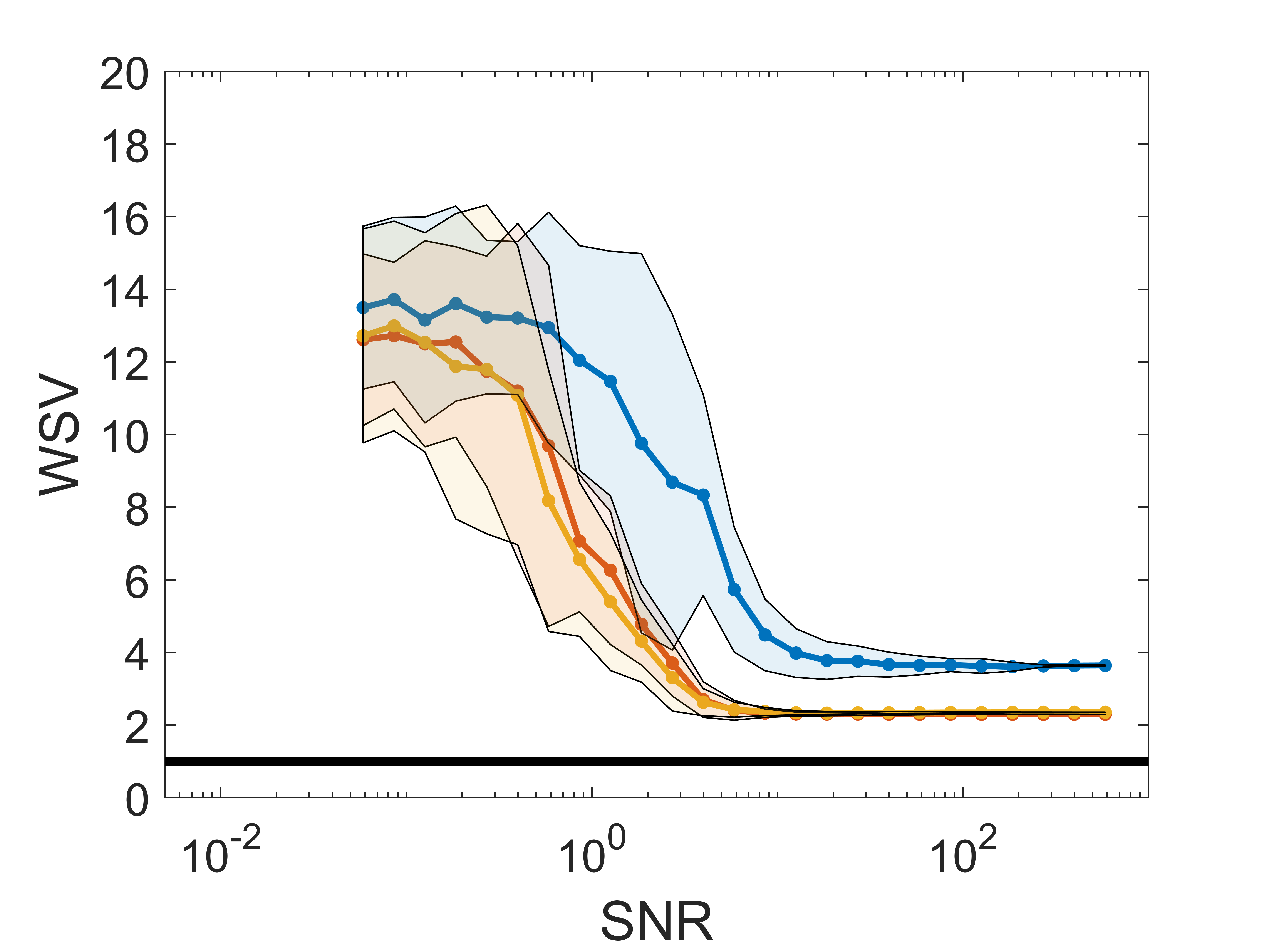}
        \captionsetup{width=0.9\linewidth}
        \caption{Perturbation 1, depth = 7.40 mm.}
        \label{fig: correlated WSV 3d simulated pert 1}
    \end{subfigure}
    \hfill
    \begin{subfigure}{0.49\textwidth}
        \centering         
        \includegraphics[width=\textwidth]{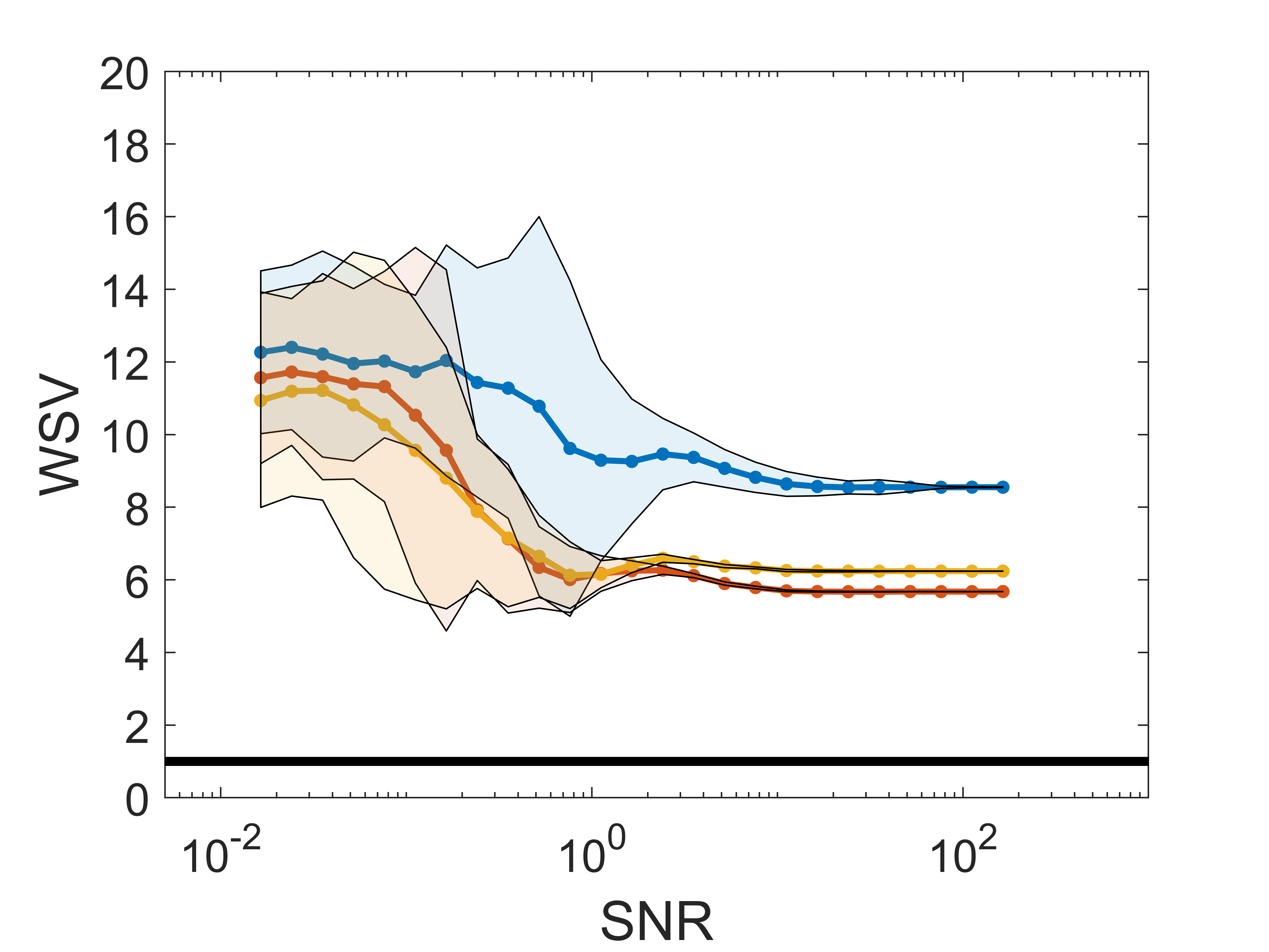}
        \captionsetup{width=0.9\linewidth}
        \caption{Perturbation 2, depth = 32.8 mm.}
        \label{fig: correlated WSV 3d simulated pert 2}
    \end{subfigure}
    \hfill
    \begin{subfigure}{0.49\textwidth}
        \centering         
        \includegraphics[width=\textwidth]{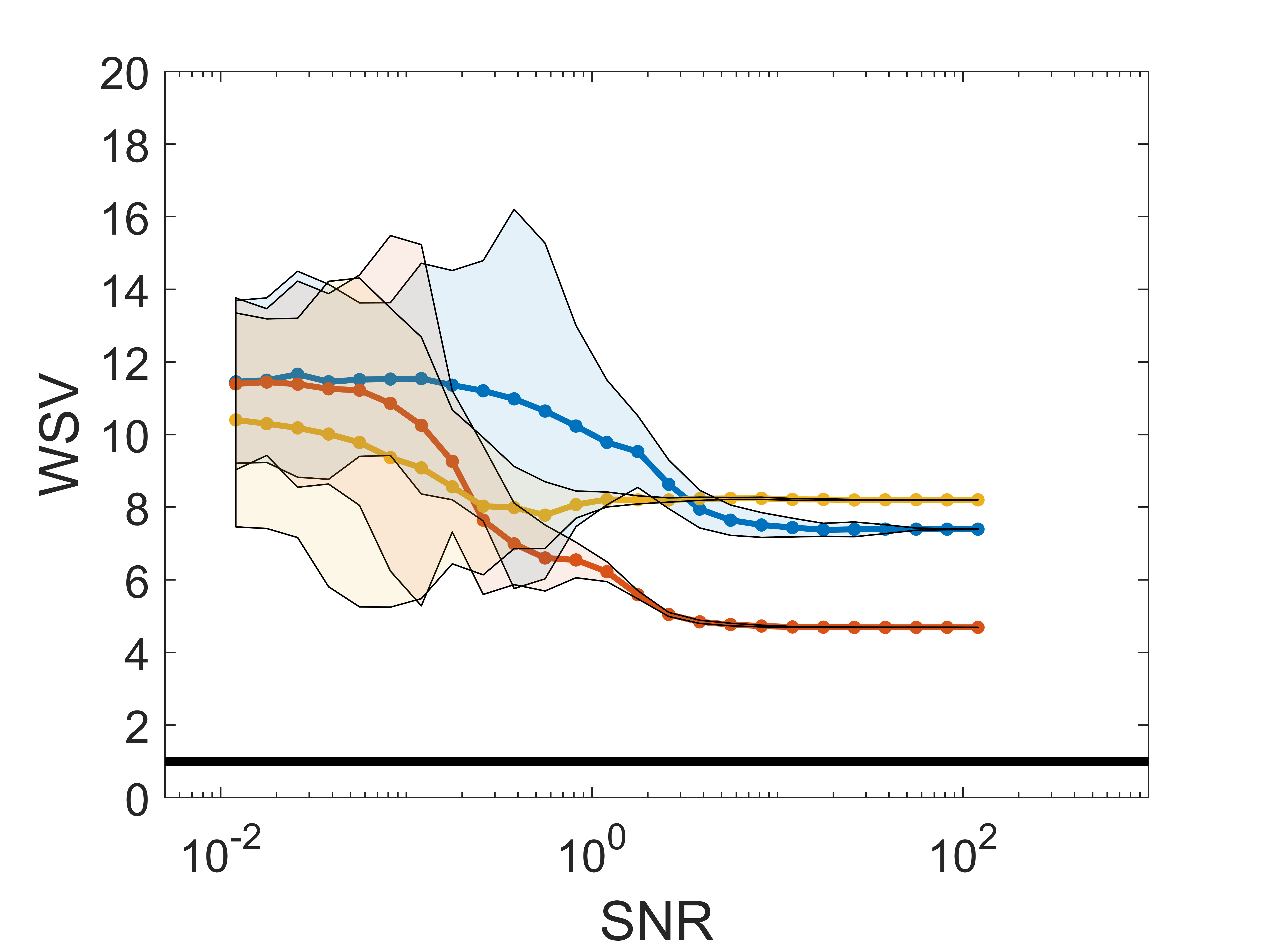}
        \captionsetup{width=0.9\linewidth}
        \caption{Perturbation 3, depth = 58.2 mm.}
        \label{fig: correlated WSV 3d simulated pert 3}
    \end{subfigure}
    \hfill
    \begin{subfigure}{0.49\textwidth}
        \centering         
        \includegraphics[width=\textwidth]{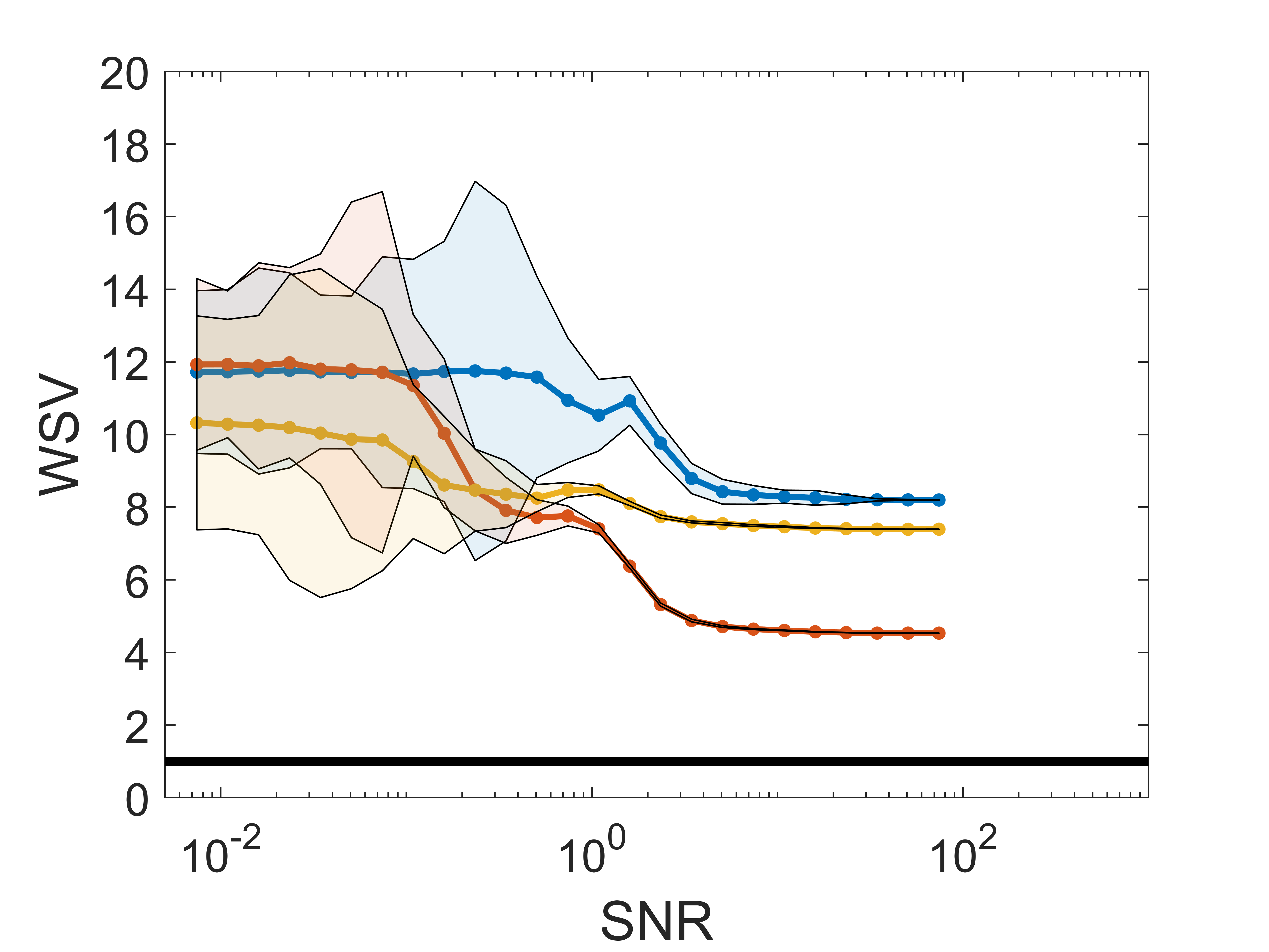}
        \captionsetup{width=0.9\linewidth}
        \caption{Perturbation 4, depth = 83.6 mm.}
        \label{fig: correlated WSV 3d simulated pert 4}
    \end{subfigure}

    \caption{The scaled WSV (mean $\pm$ IQR) for each perturbation as a function of the SNR for images reconstructed from 3D simulated data with correlated noise. The WSV is shown for images reconstructed without NBC (blue), with U-NBC (orange), with C-NBC (yellow) and for the target image (black horizontal line). A WSV value closer to the black line indicates a higher image quality.}
    \label{fig: correlated WSV 3d simulated}
\end{figure}

On visual inspection, the images reconstructed without NBC did not reconstruct images corresponding to the target image at any SNR. Images reconstructed with U-NBC and C-NBC did correlate to the target image with a fidelity that degraded as SNR decreased (Fig. \ref{fig: 3D simulated examples}). U-NBC and C-NBC improved the WSV for the majority of images in all noise cases (Table \ref{tab: WSV improvement} and Fig. \ref{fig: uncorrelated WSV 3d simulated} and \ref{fig: correlated WSV 3d simulated}).

\begin{figure*}
    \centering
    \begin{subfigure}{0.8\textwidth}
        \centering
        \includegraphics[width=0.9\textwidth]{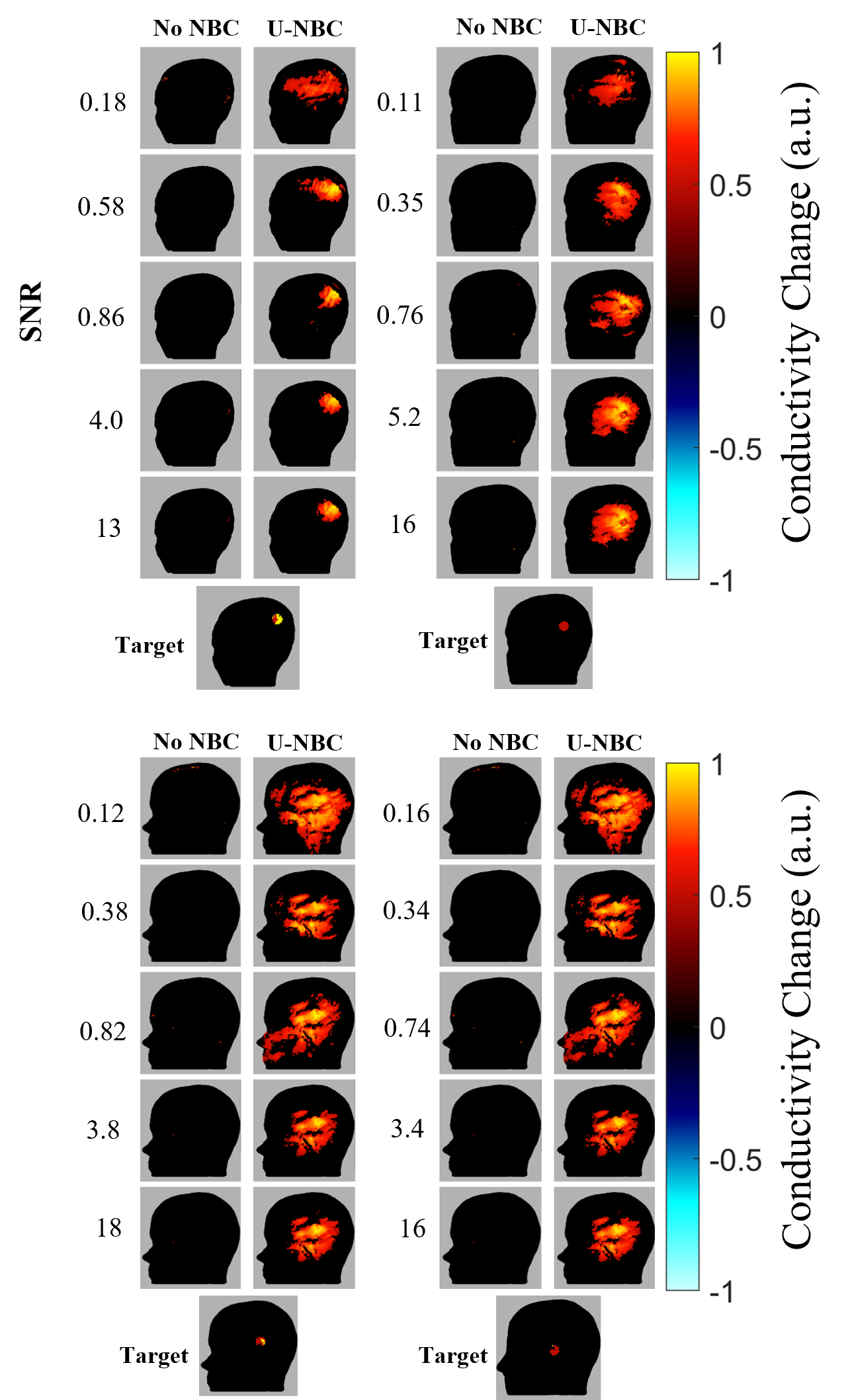}
        \caption{}
        \label{fig: uncorrelated 3D simulated example}
    \end{subfigure}
    \end{figure*}
    \begin{figure*}\ContinuedFloat
    \begin{subfigure}{\textwidth}
        \centering
        \includegraphics[width=0.97\textwidth]{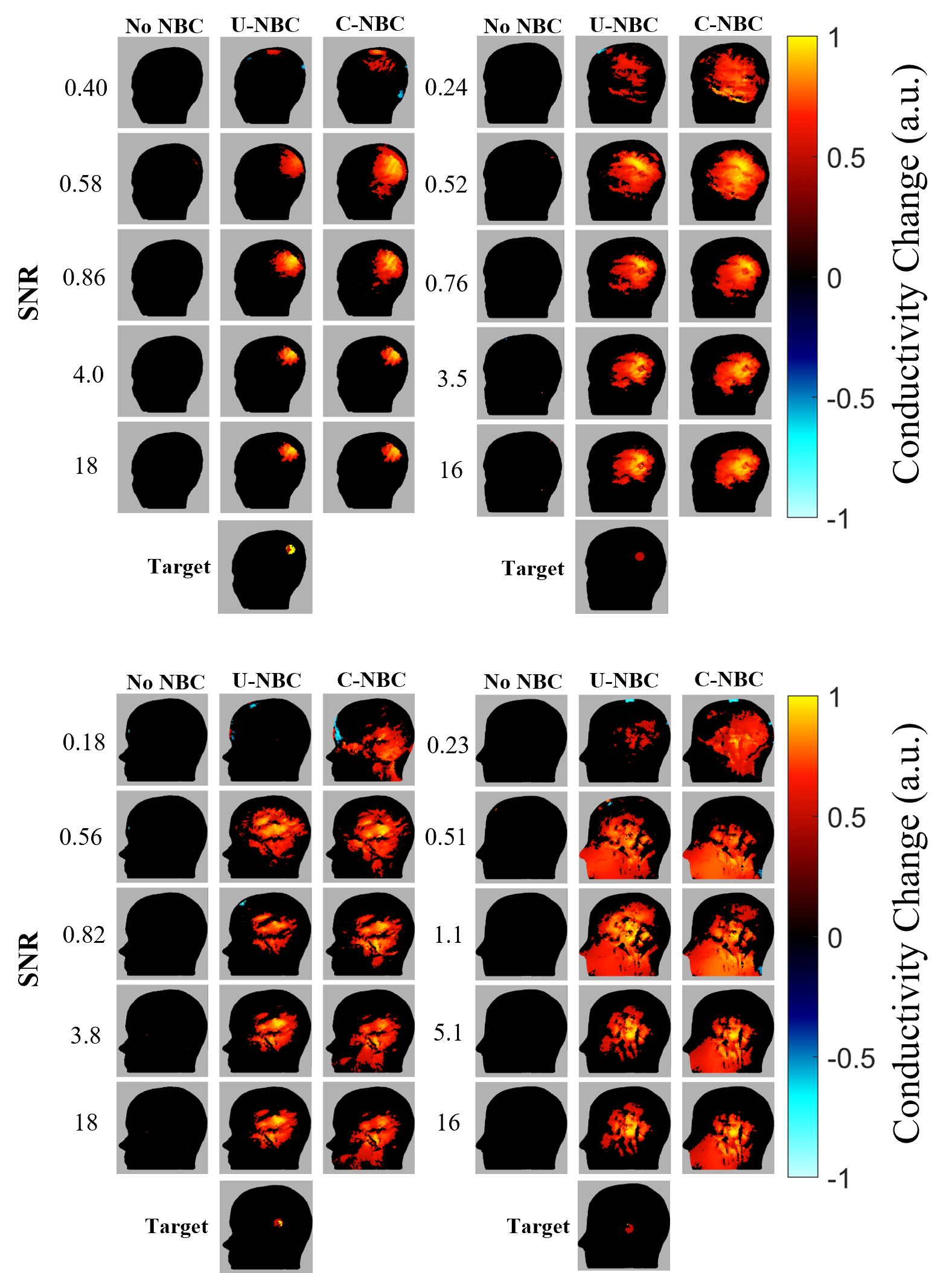}
        \caption{ }
        \label{fig: correlated 3D simulated example}
    \end{subfigure}

    \caption{Sagittal slices of example reconstructions at five SNRs of all perturbations reconstructed from simulated data in 3D with (a) uncorrelated Gaussian noise and (b) Correlated Gaussian noise. All images were thresholded at 50 \%. of the largest absolute change. Each slice is taken through the centre of mass of the true perturbation.}
    \label{fig: 3D simulated examples}
\end{figure*}

\subsection{2D Saline Tank}

\begin{figure}
    \centering
    \includegraphics[width = \textwidth]{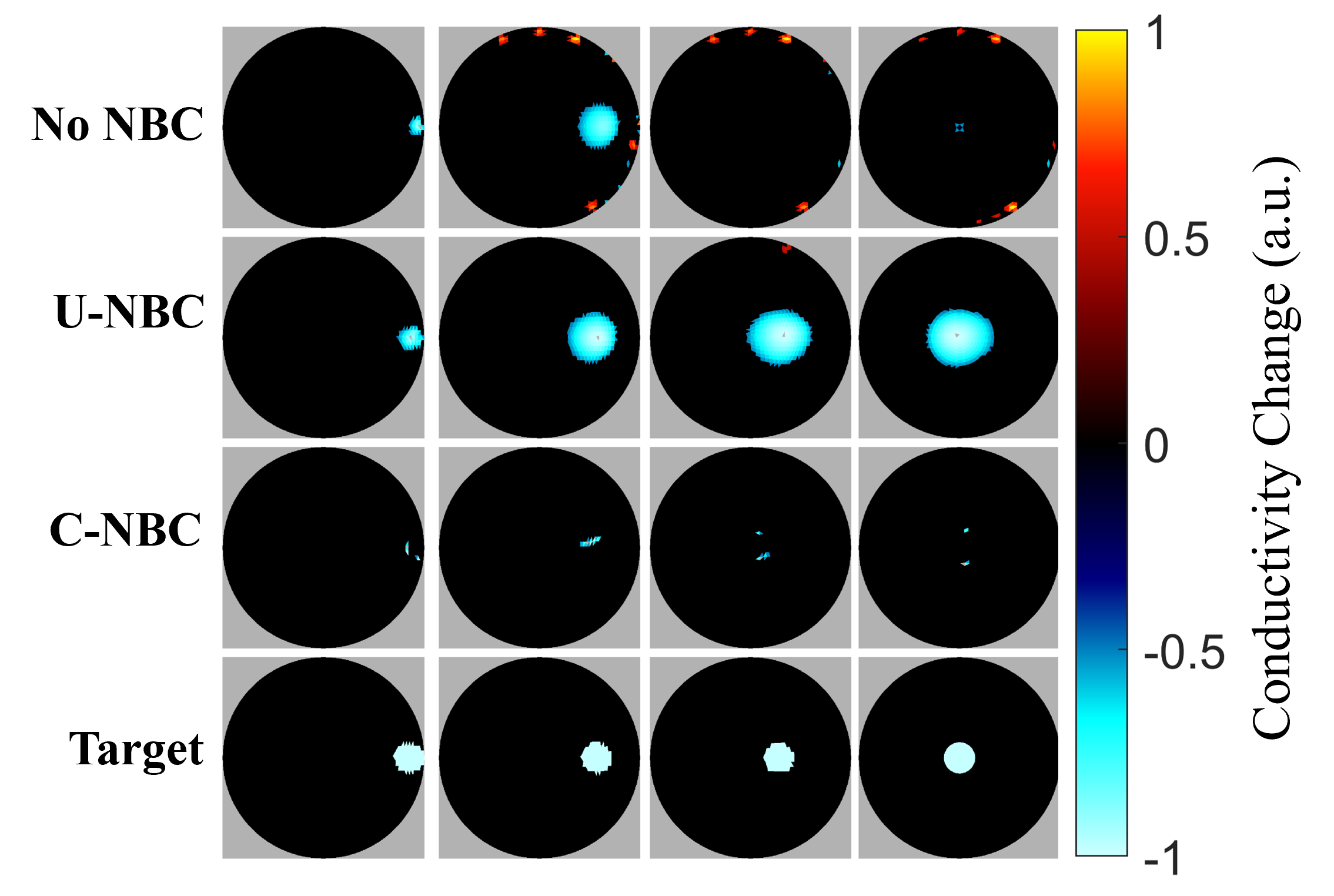}
    \caption{Example 2D image reconstructions for a perturbation in a cylindrical saline tank. The images have been thresholded at 50Fig. of the maximum absolute conductivity change. The injection protocol used for these images was `skip-2'.}
    \label{fig: 2D tank reconstructions}
\end{figure}

On visual inspection, images reconstructed with U-NBC were always of superior quality to those reconstructed without NBC or with C-NBC (Fig. \ref{fig: 2D tank reconstructions}). The perturbations reconstructed with C-NBC were in approximately the correct location but were not the correct shape and were fractured into multiple regions. The perturbations reconstructed without NBC did not correspond to the target perturbation for the deepest two perturbations but did for the two most superficial perturbations. The perturbations reconstructed with U-NBC correlated to the true perturbation in all cases shown and were of superior quality to the images reconstructed without NBC and with C-NBC. 

\begin{figure}
    \centering
    \includegraphics[width=\textwidth]{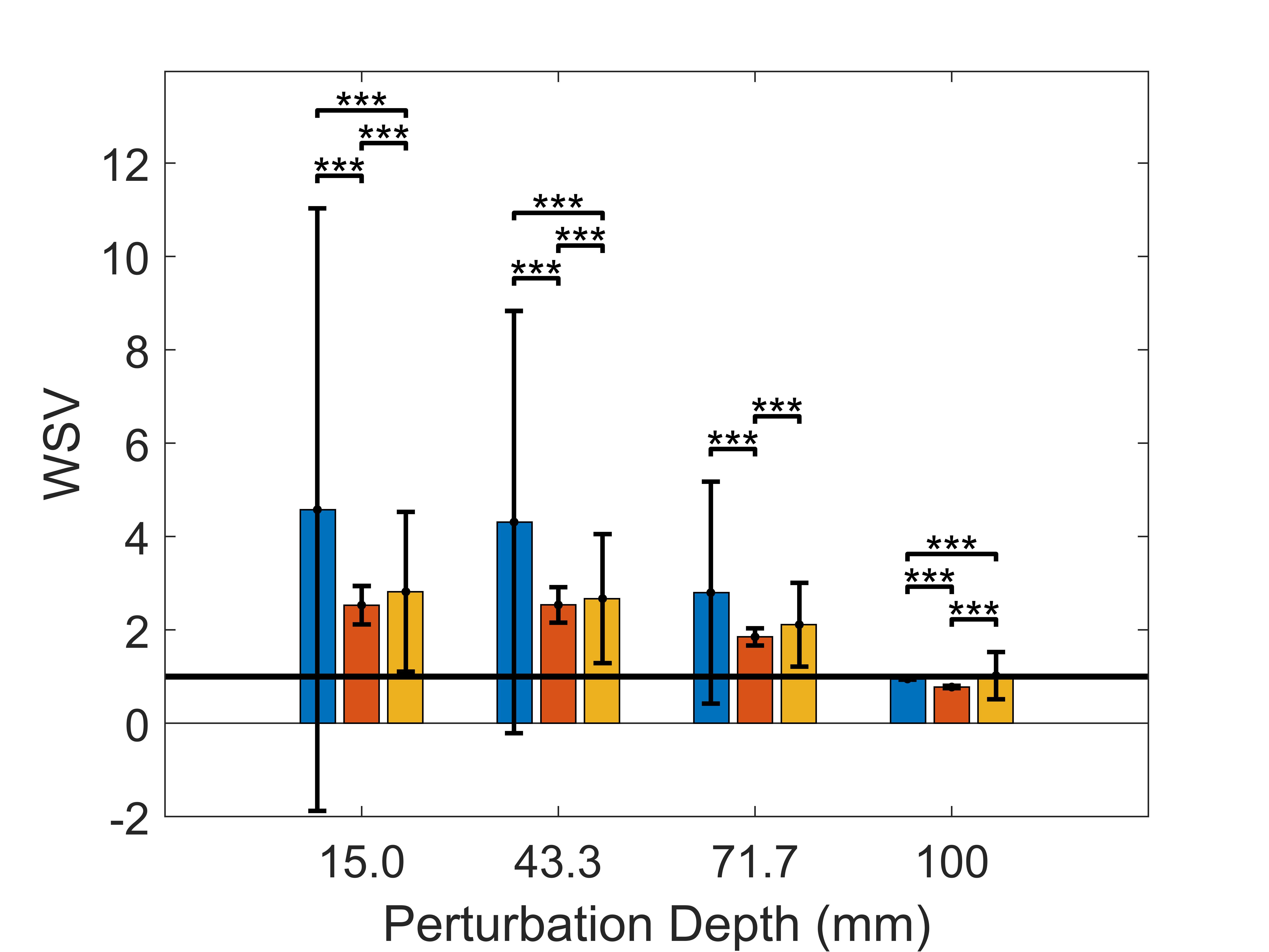}
    \caption{The WSV (mean $\pm$ IQR) for each perturbation imaged in 2D in a saline tank. The WSV is shown for the images without NBC (blue), with U-NBC (orange), with C-NBC (yellow) and the target image (black horizontal line).}
    \label{fig: 2D tank WSV}
\end{figure}

The WSV was significantly improved for the three most superficial perturbations with U-NBC and C-NBC compared to reconstructions without NBC, there was a mean difference of 0.54 and 0.22 respectively (Fig. \ref{fig: 2D tank WSV}). Without NBC, the WSV was significantly improved for the deepest perturbation with respect to U-NBC with a difference of 0.17 (Fig. \ref{fig: 2D tank WSV}). 

\subsection{3D Saline Tank}
\begin{figure}[h]
    \centering
    \includegraphics[width=\textwidth]{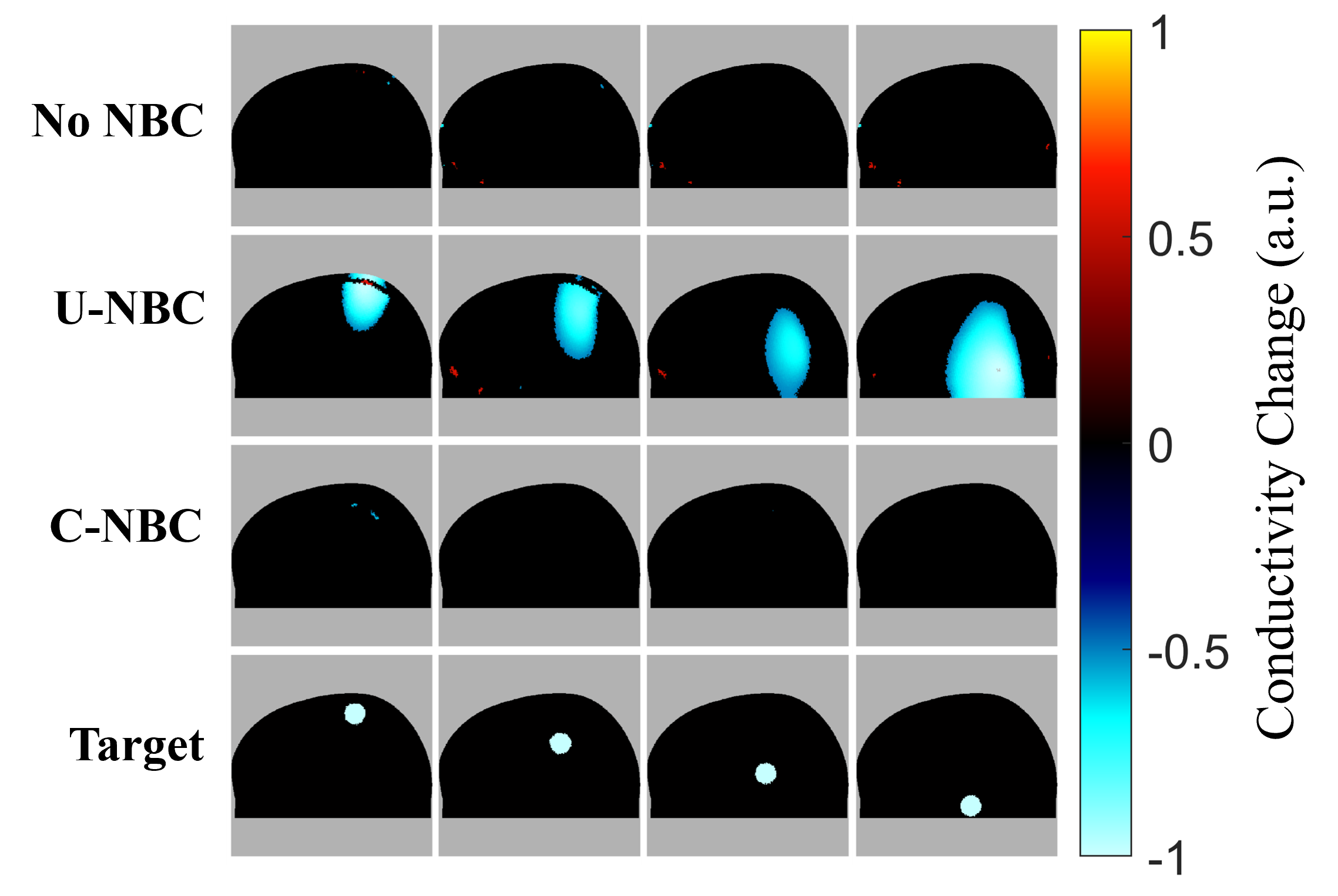}
    \caption{Example sagittal slices of image reconstructions for a perturbation in a 3D head-shaped saline tank. Each slice is taken through the centre of mass of the true perturbation. The images have been thresholded at 50 \% of the maximum absolute conductivity changes.} 
    \label{fig: head tank reconstructions}
\end{figure}

On visual inspection, images reconstructed without NBC and with C-NBC did not correspond to the target image and there was no clear perturbation reconstructed. Images with U-NBC reconstructed perturbations that corresponded to the target image (Fig. \ref{fig: head tank reconstructions}). 

\begin{figure}[h]
    \centering
    \includegraphics[width=\textwidth]{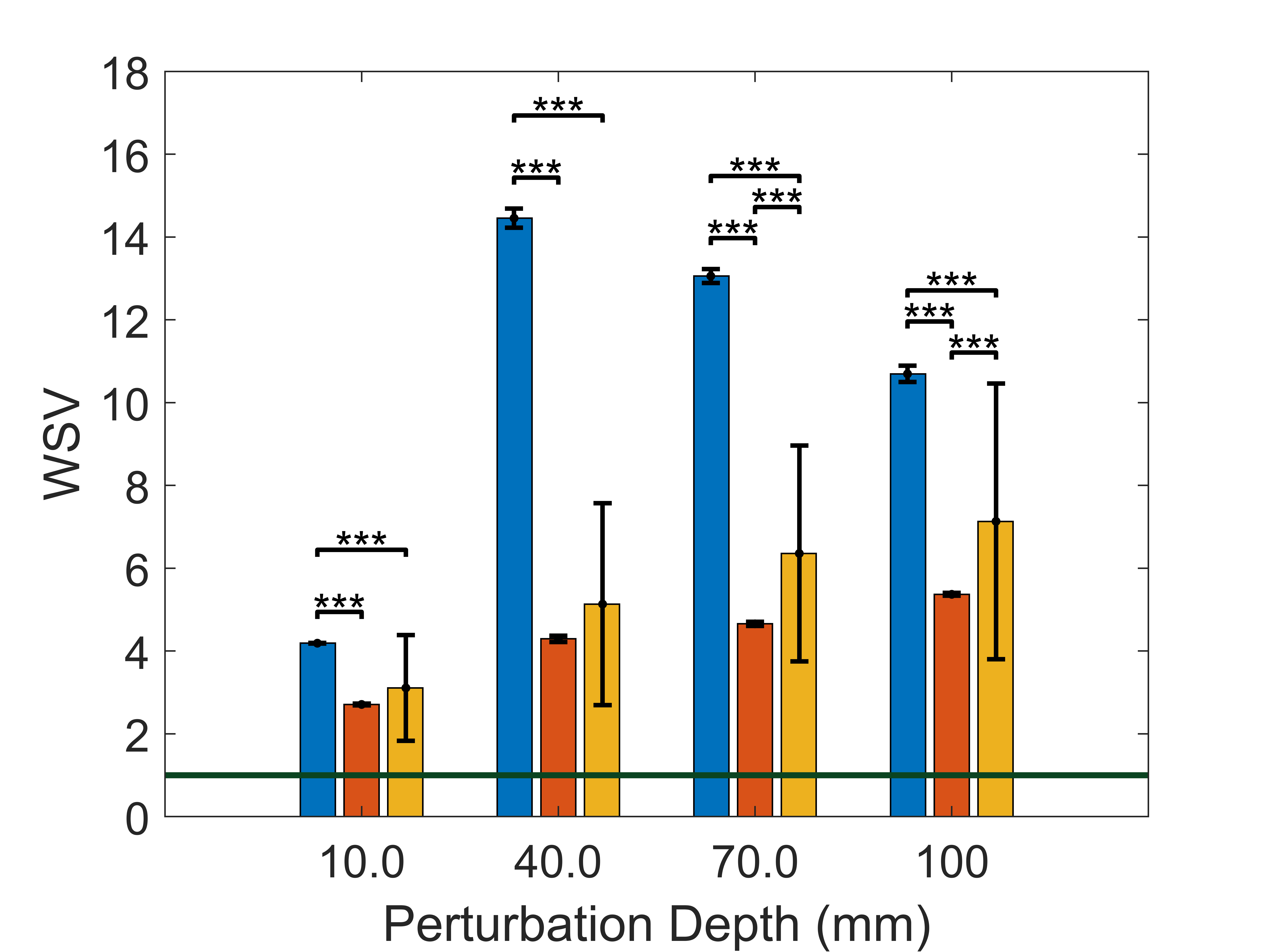}
    \caption{The WSV (median $\pm$ IQR) for each perturbation imaged in 3D in a saline tank. The WSV is shown for the images without NBC, with U-NBC, with C-NBC and the target image.}
    \label{fig: 3D tank WSV}
\end{figure}

The use of U-NBC and C-NBC significantly improved the WSV for all perturbations compared to reconstructions without NBC with a mean difference of 6.3 and 5.6 respectively. However, for C-NBC this did not correspond to an increase in quality on visual inspection (Fig. \ref{fig: 3D tank WSV}).

\subsection{\textit{In Vivo} Lung Imaging}
\begin{figure}[h]
    \centering
    \includegraphics[width = \textwidth]{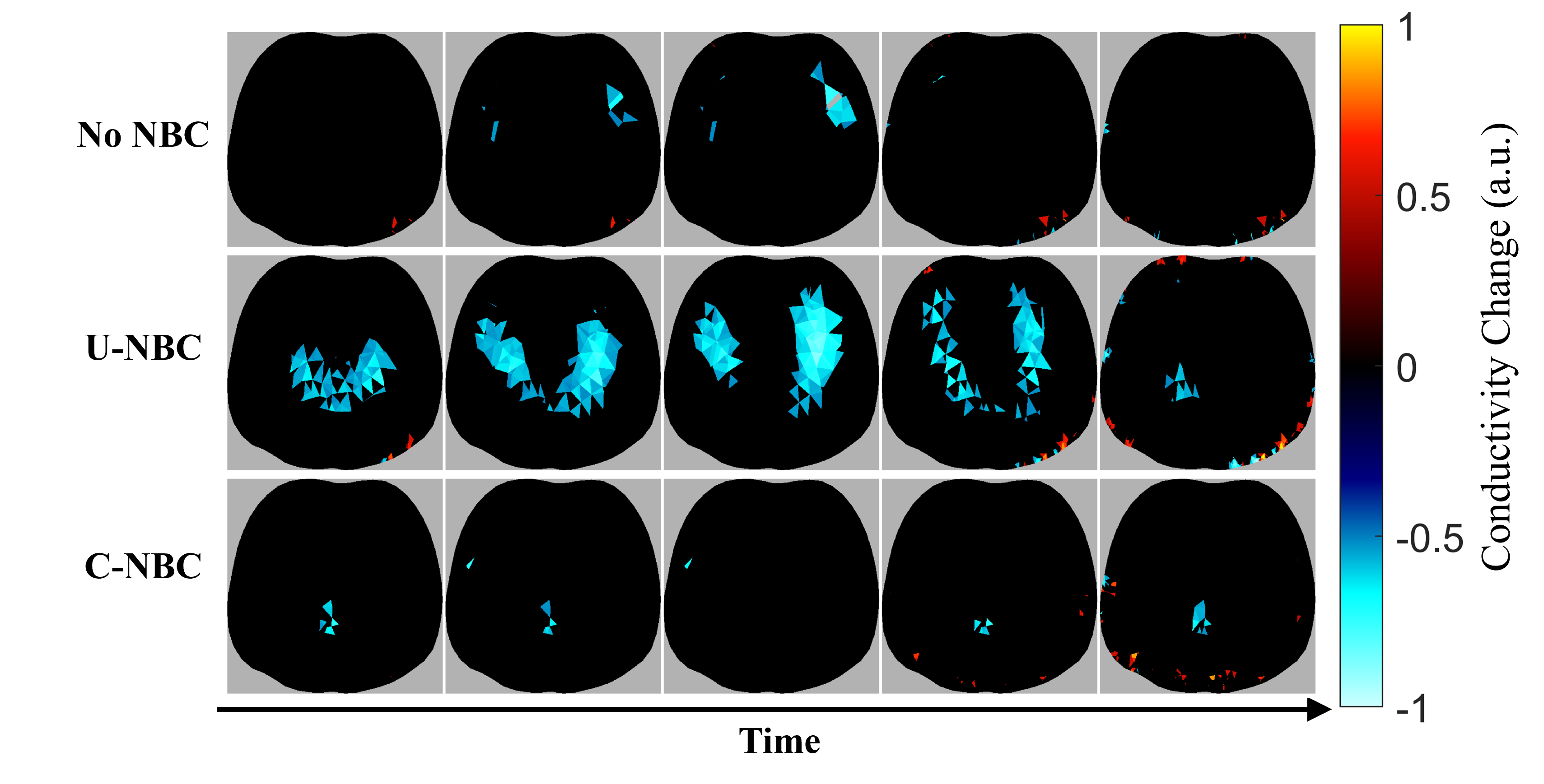}
    \caption{Transverse slices of EIT reconstructions of lung ventilation during inhalation in the thorax of a human over approximately 6 s. All slices were taken through the plane of the ring of electrodes. Reconstructions are shown without NBC, with U-NBC and with C-NBC. The images have been thresholded at 50 \% of the maximum absolute conductivity changes.}
    \label{fig: lung recons}
\end{figure}

On visual inspection, reconstructed images of lung ventilation during inhalation were of superior quality when NBC was used, with U-NBC outperforming C-NBC. The images with U-NBC most clearly show lung-shaped regions of negative conductivity change, whereas C-NBC and no NBC both show a large number of artefacts (Fig.\ref{fig: lung recons}).

\subsection{\textit{In Vivo} Fast Neural Imaging}

\begin{figure}[h]
    \centering
    \includegraphics[width = \textwidth]{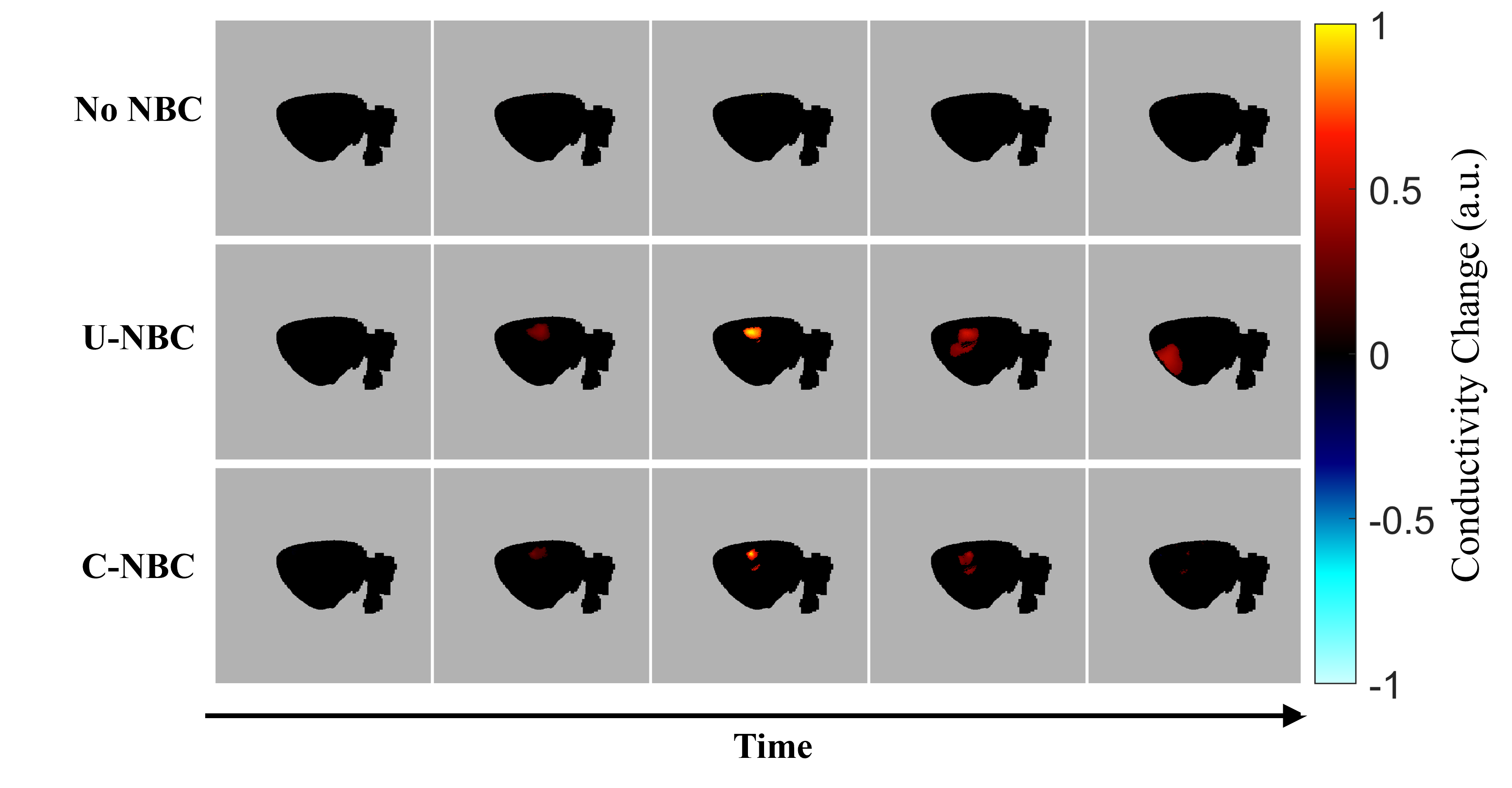}
    \caption{Sagittal slices of EIT reconstructions of evoked activity in the barrel cortex of the rat brain due to whisker stimulation over approximately 20 ms. The posterior of the brain is on the right-hand side. All slices were taken through the barrel cortex. Reconstructions are shown without NBC, with U-NBC and with C-NBC. The images have been thresholded at 50 \%. of the maximum absolute conductivity changes.}
    \label{fig: rat recons}
\end{figure}

On visual inspection, the overall image quality is superior when U-NBC or C-NBC is used compared to the case without NBC (Fig.\ref{fig: rat recons}). The case without NBC reconstructs nothing inside the barrel cortex, indicating that the largest conductivity changes are artefacts, whereas the two cases with NBC reconstruct clear perturbations corresponding to neural activity in the barrel cortex (Fig.\ref{fig: rat recons}).  

\section{Discussion}
\label{sec: discussion}

\subsection{Summary of Results}

In summary, the propagation of Gaussian noise in the signal has been shown to remain as Gaussian noise in the reconstructed image for the linearised problem of EIT and the post-processing technique of NBC has been mathematically developed to improve quality in impedance tomography images. NBC was demonstrated to significantly improve image quality for reconstructed perturbations in 2D homogeneous domains and 3D heterogeneous domains \textit{in silico} based on visual inspection of the images and calculation of the WSV. NBC was most effective at an SNR of $\sim 1$ in 2D and 3D (Fig. \ref{fig: uncorrelated WSV 2d simulated}, \ref{fig: correlated WSV 2d simulated}, \ref{fig: uncorrelated WSV 3d simulated} and \ref{fig: correlated WSV 3d simulated}). For low SNR ($< 1$), NBC consistently improved image quality in 2D and improved image quality in nearly all cases in 3D, for high SNR ($>10$) NBC had almost no change in the image quality in 2D and increased image quality in nearly all cases in 3D (Fig. \ref{fig: 2D simulated examples} and \ref{fig: 3D simulated examples}). These results were validated experimentally, with the quality of 2D and 3D image reconstructions being shown to improve with NBC (Fig. \ref{fig: 2D tank reconstructions} and \ref{fig: head tank reconstructions}). For correlated noise, C-NBC reconstructed superior images to U-NBC \textit{in silico} but inferior images in practice. For \textit{in vivo} EIT data, NBC always improved the quality of the images on visual inspection and preserved the expected shape of the perturbation (Fig.s \ref{fig: lung recons} and \ref{fig: rat recons}).

\subsection{Is NBC Mathematically Coherent?}

The preservation of Gaussian noise through the reconstruction algorithm allows the standard deviation of the reconstructed noise to serve as a metric for pixel/voxel-wise reconstructed noise amplitude in the image. This provides a solid mathematical basis for NBC as a post-processing technique to improve image quality.

\subsection{Should NBC be Used in EIT?}

This study demonstrates that the post-processing technique of NBC is a valuable tool for EIT. For low SNR signals in 2D simulations, NBC can clearly be seen to transform images from complete noise to accurate reconstructions of the considered perturbation. For medium SNR signals, NBC improved the fidelity of the imaged perturbation and for high SNR signals, NBC preserved image quality (Fig. \ref{fig: 2D simulated examples}). This was supported by a statistical analysis of the WSV (Fig. \ref{fig: uncorrelated WSV 2d simulated} and \ref{fig: correlated WSV 2d simulated}). Experimentally, U-NBC was capable of reducing noise in all reconstructions, transforming unsuccessful reconstructions from noise to faithful images and preserving the quality of successful reconstructions (Fig. \ref{fig: 2D tank reconstructions}). Even for cases where the WSV was inferior with NBC (i.e. the most superficial perturbation in Fig. \ref{fig: 2D tank reconstructions}), NBC did not decrease image quality on visual inspection.

In 3D, the images reconstructed \textit{in silico} without NBC bear no resemblance to the target image at any SNR level. It is only when NBC is applied that the images correspond to the target image by eye, this clearly shows that, in this case, NBC is necessary for achieving an accurate reconstruction. This was validated experimentally with reconstructions of a perturbation in a head-shaped tank bearing no resemblance to the target image unless U-NBC was used (Fig. \ref{fig: head tank reconstructions}).

NBC was capable of transforming noisy reconstructions to accurate images for SNR values as low as 0.005 in 2D (perturbation 4 in Fig. \ref{fig: correlated 2D simulated example}) and 0.6 in 3D (perturbation 1 in Fig. \ref{fig: uncorrelated 3D simulated example}) which shows that NBC increases EIT's robustness to measurement noise. Overall it is recommended that NBC always be applied as a post-processing step for EIT images because it has the capacity to completely transform images from noise to accurate reconstructions at low SNR, improve image quality at medium SNR and preserve high image quality at high SNR. 

\subsection{Uncorrelated or Correlated NBC?}

For 2D reconstructions \textit{in silico}, C-NBC performed better than U-NBC in terms of WSV (Fig. \ref{fig: correlated WSV 2d simulated}), this can be observed for the lower SNR cases (Fig. \ref{fig: correlated 2D simulated example}) where C-NBC reconstructs a more distinct, higher fidelity perturbation, particularly for the two deepest perturbations. For 3D reconstructions \textit{in silico}, C-NBC and U-NBC performed similarly for the two most superficial perturbations which is clear from the image reconstructions (Fig. \ref{fig: 3D simulated examples}) as well as the WSV error (Fig. \ref{fig: correlated WSV 3d simulated}). For the deeper two perturbations, the WSV error showed that U-NBC was superior to C-NBC (Fig. \ref{fig: correlated WSV 3d simulated}). However, the reconstructions for these two perturbations were of low quality in all cases. Whilst there is some correspondence between the target image and reconstructions, this would likely be deemed a failed reconstruction in practice. For this reason, these perturbations do not serve as a good comparison between C-NBC and U-NBC, meaning it is unclear from these images which was superior in practice.

For 2D and 3D experimental image reconstructions, it is immediately clear that U-NBC is superior. C-NBC did not reconstruct images which accurately corresponded to the target image in any case. An analysis of the measured noise showed that, for many channels, the measured noise distribution was non-Gaussian, which explains the failure of the C-NBC since a Gaussian noise distribution is a prerequisite of NBC. 

To summarise, C-NBC is likely preferable to U-NBC if all of the measurement noise is normally distributed. However, in practice, some non-Gaussian noise will likely be measured. In this case, U-NBC reconstructs images of superior quality. It is therefore recommended that U-NBC be used.

\subsection{Does Noise-based Correction Improve \textit{In Vivo} Images?}

The two examples of NBC applied to \textit{in vivo} data clearly demonstrate that NBC is a valuable technique for improving the image quality in a real experimental setup with physiological data. For the case of imaging lung ventilation in a human, the data was noisy and collected with only one plane of electrodes. The uncorrected images did not display any clear lung-shaped regions and contained a large presence of artefacts, the images with C-NBC slightly reduced the artefacts but did not return any lung-shaped regions. The images with U-NBC were largely improved, with clear lung-shaped regions visible which corresponded to prior expectations (Fig. \ref{fig: lung recons}). For the case of evoked activity in the barrel cortex of the rat brain, the activity was not successfully reconstructed unless NBC was used (Fig. \ref{fig: rat recons}), which was verified to correspond to the expected location, size and time duration of the activity in the original work \cite{Faulkner2019}. U-NBC reconstructed images of superior quality to C-NBC, but the improvement was not large on visual inspection (Fig. \ref{fig: rat recons})

\subsection{Is Noise-based Correction Robust for a Range of Domain and Perturbation Geometries?}

This work has considered five different computational models (one in 2D and four in 3D)  for image reconstruction, each with a unique geometry, electrode configuration and conductivity distribution. In all cases, NBC was shown to be a beneficial technique which could improve image quality, showing that the principle of NBC is robust across geometries. In addition to this, the two \textit{in vivo} cases were reconstructions of irregularly shaped physiological phenomena. In each case, the reconstructed images with NBC did not significantly alter the shape of the perturbation from prior expectations, demonstrating that NBC is capable of respecting the shape of the perturbation whilst increasing image quality. 

\subsection{Study Limitations}

Whilst NBC has clearly been shown to improve image quality for EIT, there are limitations to the method. Firstly, NBC is not a `silver bullet' capable of recovering images with any level of noise, for each system, there will be a noise threshold above which NBC will be ineffective. In addition to this, NBC is not well suited to handling severely non-Gaussian noise since this is not necessarily well characterised by the standard deviation and does not necessarily maintain its distribution when propagated through the reconstruction algorithm. 

\subsection{Study Implications}
This study included the mathematical development of NBC as a post-processing technique for the linearised inverse problem in EIT. This is a broad category of reconstruction methods; however, only Tikhonov regularisation has been considered experimentally and \textit{in silico} in this study. Since the principle of NBC is mathematically coherent for all linearised solutions, it follows that NBC will likely be beneficial for all other linearised methods, despite not being explicitly tested in this study. Similarly, this work addressed NBC for EIT; however, the working principles of other impedance imaging techniques such as magnetic detection EIT \cite{Ireland2004, Mason2023, Mason2023b}, magnetic induction tomography \cite{Watson2021} and electrical capacitance tomography \cite{Marashdeh2015} are almost mathematically identical to EIT and are all inverse problems which can be linearised. For this reason, it follows that NBC will likely be a beneficial technique for these applications as well. 

One potential method to further increase image quality is to combine NBC with other post-processing techniques such as the U-net-based method proposed by Herzberg \cite{Herzberg2023}. Since this is a method utilising neural networks, it will likely distort the propagation of noise and thus should be used after NBC. This is a promising area for future research and could result in greater tolerance to noise and artefacts in EIT images.

\subsection{Author Contributions}
\textbf{K.M.} Methodology, software, validation, formal analysis, investigation, writing, visualisation. \textbf{F.M.A.} Investigation, formal analysis, writing, visualisation. \textbf{D.H.} Supervision. \textbf{K.A.} Conceptualisation, supervision. 
 
\subsection{Data Availability}
The data and code presented here are available upon request to the corresponding author. 

\subsection{Acknowledgements}
This work was supported by the EPSRC DTP Research Studentship [EP/N509577/1 and EP/T517793/1], EPSRC Grant [EP/X018415/1] and EPSRC-funded UCL i4health CDT [EP/S021930/1].

\subsection{Conflict of Interests}
D.H. is the director of Cyqiq Ltd which has an interest in commercialising fast neural EIT.

\clearpage

\bibliography{bibliography}
\bibliographystyle{IEEEtran}

\end{document}